\documentclass{aa}
\usepackage{graphicx}

\def\psfig#1{{}}	

\def\psfig#1#2\includegraphics{{}}

\def\parn{\par\noindent}

\def\los{line of sight}

\def\R#1{{\mathrm{#1}}}		
\def\Sec#1{{Section~\ref{s:#1}}}
\def\Eq#1{{Eq.~\ref{e:#1}}}	

\def\EQN#1{\label{e:#1}}        
\def\Fig#1{{Fig.~\ref{f:#1}}}	

\def\M#1{{\mathbf{#1}}}	
\def\T#1{{{#1}^{\bot}}}		
\def\d#1{{\R{d}{#1}}}		

\def\Mfunction#1{{\rm #1}}
\def\Mvariable#1{{\rm #1}}
\def\Muserfunction#1{{\rm #1}}

\def\ScriptCapitalA{{\cal A}}

\begin{document}

\title{Physical properties and small-scale structure
of the Lyman-$\alpha$ forest: Inversion of the HE~1122$-$1628 UVES 
spectrum\thanks{Based on data collected during Science Verification
of the Ultra-violet and Visible Echelle Spectrograph at
the European Southern Observatory on the 8.2~m KUEYEN telescope
operated on Cerro Paranal, Chile.}\thanks{Table A.1 is only available 
at the CDS via anonymous ftp to cdsarc.u-strasbg.fr}}
\titlerunning{Inversion of the HE~1122$-$1628 UVES 
spectrum}

  \author{ Emmanuel~Rollinde \inst{1}, Patrick~Petitjean \inst{1,2}, 
Christophe~Pichon \inst{3,1}}

   \offprints{rollinde@iap.fr}

\institute{$^1$ Institut d'Astrophysique de Paris, 98 bis boulevard
       d'Arago, 75014 Paris, France \\
       $^2$ UA CNRS 173 -- DAEC, Observatoire de Paris-Meudon, F-92195 Meudon
Cedex, France \\
$^3$ Observatoire de Strasbourg, 11 rue de l'Universit\'e,
67000 Strasbourg, France \\
}

\date{Received 19 February 2001 / Accepted 31 May 2001}

\abstract{
We study the physical properties of the Lyman-$\alpha$ forest
by applying the 
inversion method 
described by Pichon et al. (2001)
to the high resolution and high S/N ratio spectrum of the $z_{\rm em}$~=~2.40 
quasar HE~1122$-$1628 obtained during Science Verification of UVES at the VLT.
We compare the column densities obtained with the new fitting procedure 
with those derived using standard Voigt profile methods. The agreement is  
good and gives confidence in the new description of the Lyman-$\alpha$ forest 
as a continuous field as derived from our method. We show that the 
observed number density of lines with log~$N$~$>$~13 and 14 is, 
respectively, 50 and 250  per unit redshift at $z$~$\sim$~2. 
We study the physical state of the gas, neglecting peculiar 
velocities, 
 assuming a relation between
the overdensity and the temperature. $T$~=~${\overline T}(\rho(x)/{\bar \rho})^{2\beta}$.
There is an intrinsic degeneracy between the parameters $\beta$ and
${\overline T}$.
We demonstrate that, at a fixed $\beta$, the temperature
at mean density, ${\overline T}$, can be uniquely extracted,
however.
While applying the method to HE~1122$-$1628, we conclude that
for 0.2~$<$~$\beta$~$<$~0.3, 
6000~$<$~${\overline T}$~$<$~15000~K at $z$~$\sim$~2.
We investigate the small-scale structure of strong absorption lines 
using the information derived from the Lyman-$\beta$, Lyman-$\gamma$ and 
C~{\sc iv} profiles. Introducing the Lyman-$\beta$ line 
in the fit allows us to reconstruct the density field up to 
$\rho/{\bar \rho}$~$\sim$~10 instead of 5 for the Lyman-$\alpha$ line only. 
The neutral hydrogen density is of the order of
$\sim$2$\times$10$^{-9}$~cm$^{-3}$ and the C~{\sc iv}/H~{\sc i} ratio varies 
from about 0.001 to 0.01 within the complexes of total column density
$N$(H~{\sc i})~$\sim$~10$^{15}$~cm$^{-2}$. Such numbers are expected 
 for photo-ionized gas of density 
$n_{\rm H}$~$\sim$~10$^{-4}$~cm$^{-3}$ and [C/H]~$\sim$~$-$2.5. There may be
small velocity shifts ($\sim$10~km~s$^{-1}$) between the peaks in the 
C~{\sc iv} and H~{\sc i} density profiles. Although the statistics is small, 
it seems that C~{\sc iv}/H~{\sc i} and $n_{\rm HI}$ are {\sl anti-correlated}.
This could be a consequence of  the high 
sensitivity of the C~{\sc iv}/H~{\sc i} 
ratio to temperature. The presence of associated 
O~{\sc vi} absorption, with a similar profile, confirms that the gas 
is photo-ionized and at a temperature of $T$~$\sim$~10$^5$~K. 
}
%

\maketitle

\section{Introduction}
\label{s:introduction}

The numerous absorption lines seen in the spectra of distant quasars
(the so-called Lyman-$\alpha$ forest) reveal the intergalactic medium
(IGM) up to redshifts larger than 5.  It is believed that the space
distribution of the gas traces the potential wells of the dark matter. 
Indeed, recent numerical $N$-body simulations have been successful at
reproducing the observed characteristics of the Lyman-$\alpha$ forest
(Cen et al.  1994, Petitjean et al.  1995, Hernquist et al.  1996,
Zhang et al.  1995, M\"ucket et al.  1996, Miralda-Escud\'e et al. 
1996, Bond \& Wadsley 1998).  The IGM is therefore seen as a smooth
pervasive medium which can be used to study the spatial distribution
of the mass on scales larger than the Jeans' length.  This idea is
reinforced by observations of multiple lines of sight.  It is observed
that the Lyman-$\alpha$ forest is fairly homogeneous on scales smaller
than 100~kpc (Smette et al.  1995, Impey et al.  1996) and highly
correlated on scales up to one megaparsec (Dinshaw et al.  1995, Fang
et al.  1996, Petitjean et al.  1998, Crotts \& Fang 1998, D'Odorico
et al.  1998, Young et al.  2001).  The number of known 
suitable multiple
lines of sight is small, however, and the sample need to be
significantly enlarged before any firm conclusion can be drawn.
\par\noindent
The standard method to analyze the observed Lyman-$\alpha$ forest
is to fit the spectra as a superposition of Voigt profiles (e.g.
Carswell et al. 1987). 
The IGM is then described as a juxtaposition of discrete clouds 
rather than a continuous field.
Very recently, new methods have been implemented to recover
the real space density distribution of the IGM by inversion of the 
Lyman-$\alpha$ forest. Nusser \& Haehnelt (1999, 2000) proposed an iterative
scheme based on Lucy's deconvolution method (Lucy 1974). Pichon et al.
(2001) used a general non-linear explicit Bayesian deconvolution method 
which offers the possibility to add as much a priori information 
as is available. The method has been implemented in order 
 to recover the
3D topology of the large-scale structures when applied to
the inversion of a network of adjacent lines of sight. 
\par\noindent
Here we apply this method to 
invert a high S/N ratio and high resolution spectrum of HE~1122$-$1628 
obtained with UVES at ESO and therefore investigate the physical conditions of the 
IGM at $z$~$\sim$~2. Data are presented in \Sec{data}. 
The assumed model for the
  Lyman absorption is described in \Sec{method}. The method is sketched 
in Appendix B.
In \Sec{eos} we investigate how the method can be used 
to constrain the temperature-density relation in
 the IGM and apply it to HE~1122$-$1628. \Sec{VPFIT} is dedicated to a 
comparison between a traditionnal Voigt profile decomposition 
and our method. 
In \Sec{structure} we use the additional information provided by 
higher Lyman 
series absorption profiles to deconvolve the saturated 
Lyman-$\alpha$ lines.
We use this information to investigate the C~{\sc iv}/H~{\sc i} ratio
through high density profiles~.
We finally draw our conclusions in \Sec{conclusion}

%
\section{Data}
\label{s:data}
The $m_{\rm V}$~=~15.5 and $z_{\rm em}$~=~2.40 quasar HE~1122$-$1628 was
observed during Science Verification of the Ultra-violet and Visible
Echelle Spectrograph mounted on the 8.2~m Kuyen telescope of the European 
Southern Observatory, mount Paranal (Chile). The spectrum, 
made public by ESO, was reduced and normalized by C\'edric Ledoux using 
MIDAS, the ESO data reduction package and following the pipeline step by step.
Metals were identified 
and flagged when redshifted in the Lyman-$\alpha$ forest before any
fitting procedure was applied.
Table~A.1 gives the results (position, Doppler parameter
and column densities) of fitting the Lyman-$\alpha$ forest
of HE~1122-1622 with two methods:
(i) an automatic Voigt profile procedure (Carswell et al. 1987) 
and (ii) the Bayesian inversion method (see below and 
Pichon et al. 2001 for details). The redshift  range probed by 
 the spectrum is $1.86-2.38$  for the Lyman-$\alpha$ forest
and $2.-2.38$ for the Lyman-$\alpha$ and Lyman-$\beta$ forests.

\parn
In the course of the study, we use three synthetic spectra, F$_{1}$, F$_{2}$
and F$_{3}$. The spectra were generated analytically applying the 
procedure described by Bi \& Davidsen (1997) and using 
a temperature-density relation (see \Eq{eos})
with $\beta=0.25$ and $\overline{T}$~=~10\,000, 20\,000 and 30\,000~K, 
respectively (see Choudhury et al. 2001). Photon and pixel noise 
is added in such a way that the signal-to-noise ratio is 
50, corresponding approximately to the quality of the observed spectrum.

%
\section{The Bayesian inversion method}
\label{s:method}

\begin{figure}
\includegraphics[width=8.5cm]{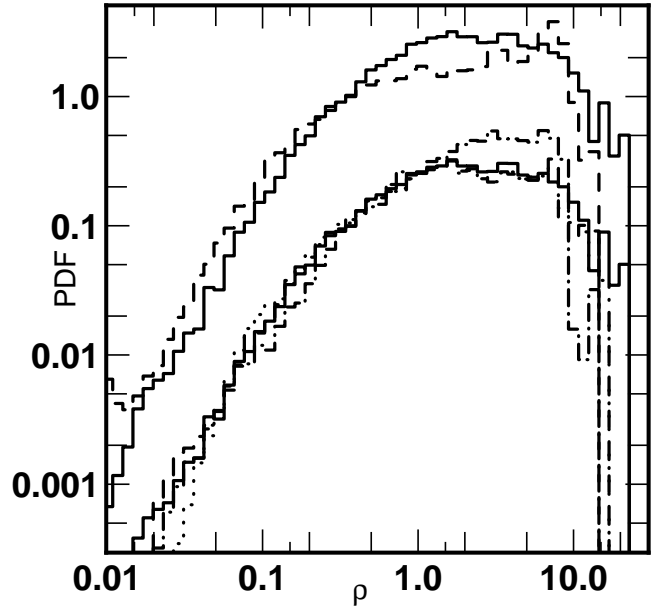}
\caption{Simulated and reconstructed probability Distribution Function (PDF), 
$\rho\ P(\log(\rho))$, of the over-density $\rho / {\bar \rho}$.
 The simulated density PDF  is plotted as a solid line.
From this density, a spectrum has been generated in real space and 
thermally broadened. 
Different amounts of noise have been added. 
The density field is then recovered using the inversion technique described 
in \Sec{method}. The dotted and dashed lines correspond to the PDF of the 
density recovered from spectra with a S/N ratio of 50
and 1000, respectively.
 The two PDFs are approximately the same except 
for the low density region.
Noise induces a lower and an upper cutoff in the recovered PDF. 
The PDF 
in the redshift-space spectrum is shown as
a dot-dashed line. The PDF recovered from the UVES spectrum
scaled up by a factor of 10 for clarity, is shown as 
another dashed line. The density PDF 
in the simulation is plotted with the same scaling, upper solid line, to aid
the comparison.}
\label{f:pdf_noise}
\end{figure}

We inverse the observed Lyman-$\alpha$ forest for the density field
using a  Bayesian method. This method is a generalisation of Wiener filtering, 
which is capable of dealing with a non-linear model for the Lyman-$\alpha$ 
absorption, including  for instance the effects of 
 gas temperature and of  peculiar motions. In this study, we assume 
the model described below, in which peculiar motions are neglected.
\parn 
The optical depth, $\tau$, in the Lyman-$\alpha$ forest is
\begin{equation}
\tau(w)= \frac{c \, \sigma_0 }{H(\overline{z}) \sqrt{\pi}}
\int_{-\infty}^{+\infty} \frac{n_{{\rm HI}}(x)}{b(x)} \exp\left(- \frac{(w-
x)^2}{b(x)^2} \right) \d x \,, \EQN{eqfun}
\end{equation} 
\parn 
where 
$c$ is the velocity of light, $\sigma_0$ is the Lyman-$\alpha$ absorption
cross section, $H$ is the Hubble constant at redshift $\overline{z}$,
$n_{{\rm HI}}$ the H~{\sc i} density field and $b$ the Doppler parameter. 
$\omega$  and $x$ are velocities (in km~s$^{-1}$). The observed flux, $F$, 
is simply $F=\exp(-\tau)$. We use $H_0$~=~75~km/s/Mpc throughout the paper.
\parn
Following Hui \& Gnedin (1997), we assume a temperature-density
 relation
\begin{equation}
T=\overline{T} \left( \frac{\rho_{\rm DM}(\M{x})}
{\bar{\rho}_{\rm DM}}\right)^{\gamma-1}\,, \EQN{eos}
\end{equation}
 where $\overline{T}$ is the 
temperature at mean dark matter density, $\bar{\rho}_{\rm DM}$, and 
$\gamma-1$ is the polytropic index. And therefore,
\begin{eqnarray}
\lefteqn{\ \ b(\M{x})=13\, {\rm km s^{-1}}\ \sqrt{\frac{\overline{T}}{10^4K}} 
\left( \frac{\rho_{\rm DM}(\M{x})}
{\bar{\rho}_{\rm DM}}\right)^{\beta}  \,,}\EQN{trac}
\end{eqnarray} 
\parn 
where $\gamma-1=2\beta$.
\parn A simple functional relation between the 
dark matter density $\rho_{\rm DM}$ and the neutral hydrogen particle density
$n_{\rm HI}$ is assumed,
\begin{eqnarray}
\lefteqn{\ \ n_{\rm    HI}(\M{x})=\overline{n}_{\rm  HI}  
 \left(  \frac{\rho_{\rm DM}(\M{x})}
{\bar{\rho}_{\rm DM}}\right)^{\alpha} \,,} \EQN{trac2} 
\end{eqnarray} 
\parn where $\alpha=2 \mbox{--} 1.4 \times\beta$.
\Eq{eqfun} reads
\begin{eqnarray}
  \tau(w)&=&A(\overline{z})c_1  \int_{-\infty}^{+\infty}
\left(
\frac{\rho_{\rm DM}(x)}
{\bar{\rho}_{\rm DM}}
\right)^{\alpha-\beta} \nonumber \\
&&\ \ \times \ \exp\left(- c_2 \frac{(w-
x)^2}{\left[{\rho_{\rm DM}(x)}
/{\bar{\rho}_{\rm DM}}
\right]^{2\beta}}  \right)  \d  x \,. \EQN{sp1}
\end{eqnarray}
The parameters $c_1$, $c_2$ and $A(\overline{z})$ depends on the
characteristic temperature of the IGM
\begin{eqnarray}
\lefteqn{\ \  c_1=
\left(13\sqrt{\pi} \sqrt{\frac{\overline{T}}{10^4}}\right)^{-1},
 \quad c_2= \left(13^2 \frac{\overline{T}}{10^4} \right)^{-1}} \nonumber \\
&\mathrm{and\ \ } A(\overline{z})=  \frac{ \overline{n}_{\rm HI}\,c \, 
\sigma_0}{H(\overline{z})} \propto \ \frac{{\overline{T}}^{-0.7}}{J} \,,
  \EQN{defc1}
\end{eqnarray}
\par\noindent 
where $J$ is the ionizing flux, assumed to be uniform. $A(\overline{z})$ is 
chosen in order to match the observed average optical depth of the
Lyman-$\alpha$ forest ($\simeq$ 0.2 at 
$z$~=~2). In the case of HE~1122$-$1628, we measure 
$\ScriptCapitalA(\overline{z})=0.22$ in the UVES spectrum and therefore use 
this value in the following.
\parn 
We   assume  that   the  a priori density  obeys   a  lognormal   distribution
characterised  by   its  a  priori  correlation   matrix. 
We plot in Fig.~\ref{f:pdf_noise} the PDF of the 
analytical density field used to 
generate spectrum F$_1$ (solid line; the peculiar velocities are not
considered) together with the PDF reconstructed after inversion of F$_1$ 
(dashed and dotted lines for S/N ratios of 50 and 1000 respectively). 
Overplotted as a dashed line, but shifted by a factor of ten for clarity,
 is the PDF for the HE~1122$-$168
spectrum. This justifies the lognormal approximation. 
The discrepancy beyond $\rho / {\bar \rho}$~$\sim$~1 is due to 
redshift distortion (see Nusser et al., 2000).
\parn The  sought
parameters, $\M{M}$,  are  the  logarithm  of the  continuous  field 
 $\rho_{\rm
DM}/\bar{\rho}_{\rm  DM}$ (named $p$ in the following),
  together  with the two
parameters of \Eq{eos}, $\overline{T}$  and $\beta$.
The priors of the inversion are described by the above 
equations~\ref{e:eos},~\ref{e:trac},~\ref{e:trac2}, plus a probability
distribution of $\M{M}$, involving a prior guess $\M{M}_0$
and a correlation function $\M{C}_0$ for $p$
\begin{equation}
 \M{C}_{0_{i,j}}=\sigma_p\  \times\ \exp\left(-\frac{(x_i-x_j)^2}{\xi_r^2}\right)\,.
 \end{equation}
In this work, $\M{M}_0\equiv \left(p_0=0,\ \overline{T} 
{\rm \ and\ } \beta {\rm \ fixed}\right)$. 
The model that relates the data $\M{D}$ and the
parameters $\M{M}$ is
 given by \Eq{sp1}.

\begin{figure*}
\includegraphics[width=12cm,angle=-90]{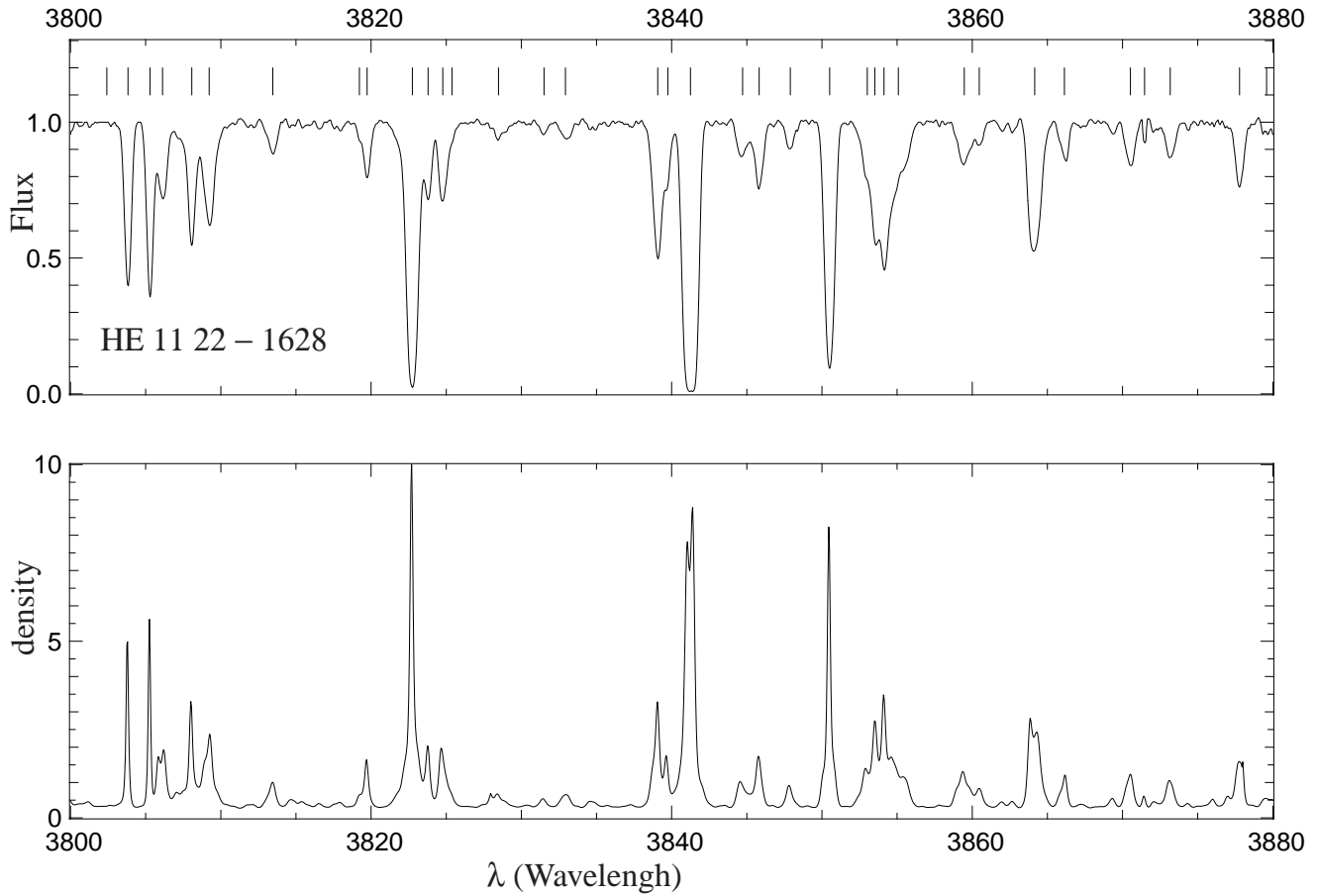}
\caption{Portion of the spectrum of HE~1122$-$1628 (upper panel) and 
corresponding  reconstructed DM density (bottom panel) when peculiar velocities are not
considered. The vertical marks indicate the positions of individual components 
found by the automatic Voigt profile fitting procedure. The S/N ratio in the
data is of the order of 60 per pixel.}
\label{f:spectre}
\end{figure*}

%
%
\parn
An example of 
the reconstruction is given in Fig.~\ref{f:spectre} which shows 
part of the 
spectrum of HE~1122$-$1628 (upper panel) and the reconstructed density 
(lower panel).
%
%
%

\section{Constraining the temperature at mean density}
\label{s:eos}

%
The method described in Section~\ref{s:method} is applied 
to the spectrum of HE~1122$-$1628 to constrain the parameters of the
temperature-density relation : the exponent $\beta$ and
the temperature at the mean density  $\overline{T}$ (see \Eq{eos}). 
The inversion is performed for a grid of $\beta$ and $\overline{T}$ values. 
Following Hui \& Gnedin (1997), we restrict our study to $\beta$ in the range
0.2--0.3.
%
\begin{figure}
\includegraphics[width=8.5cm]{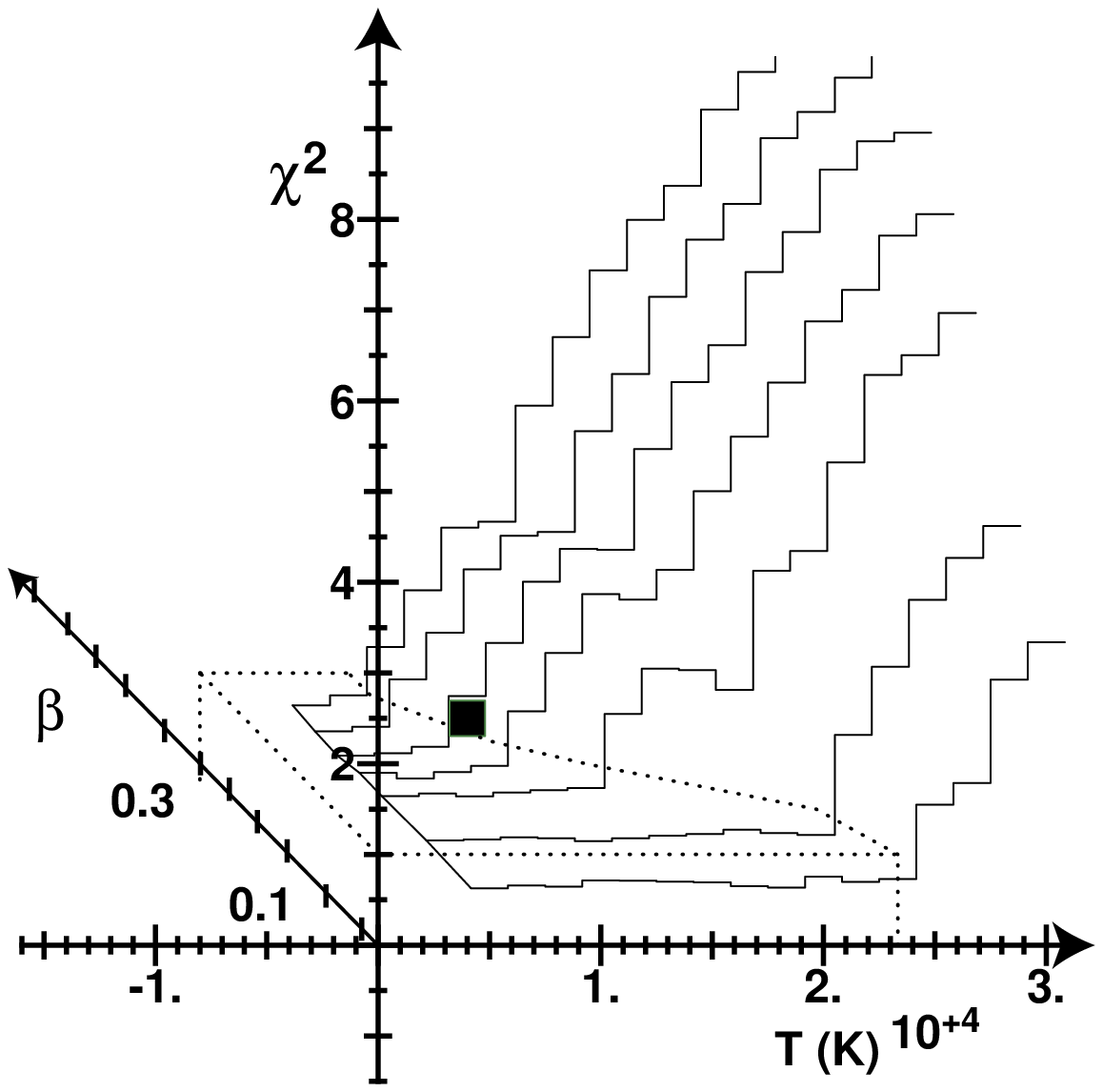}
\caption{Result of the fit  of a synthetic spectrum  
constructed from an analytical simulation and assuming an temperature-density relation defined by \Eq{eos} with $\beta_0$~=~0.25 and $\overline{T}_0$~=~10\,000 K
(spectrum F$_1$, see \Sec{data}). The inversion is performed for 
the density field, for a grid of couples 
($\beta$,$\overline{T}$), and the final minimum reduced $\chi^2$ is plotted 
versus $\beta$ and $\overline{T}$. The plane corresponding to $\chi^2=1$ is 
shown with dotted lines.
The locus of the points ($\beta$,$\overline{T}_{}$,$\chi^2$=1) is what
 is refered to as the borderline in the text. 
The correct couple ($\beta_0$, $\overline{T}_0$),
marked by a filled square, is located  on  this borderline. 
}
  \label{f:chi_plane}
\end{figure}
\begin{figure}
\includegraphics[width=8.cm]{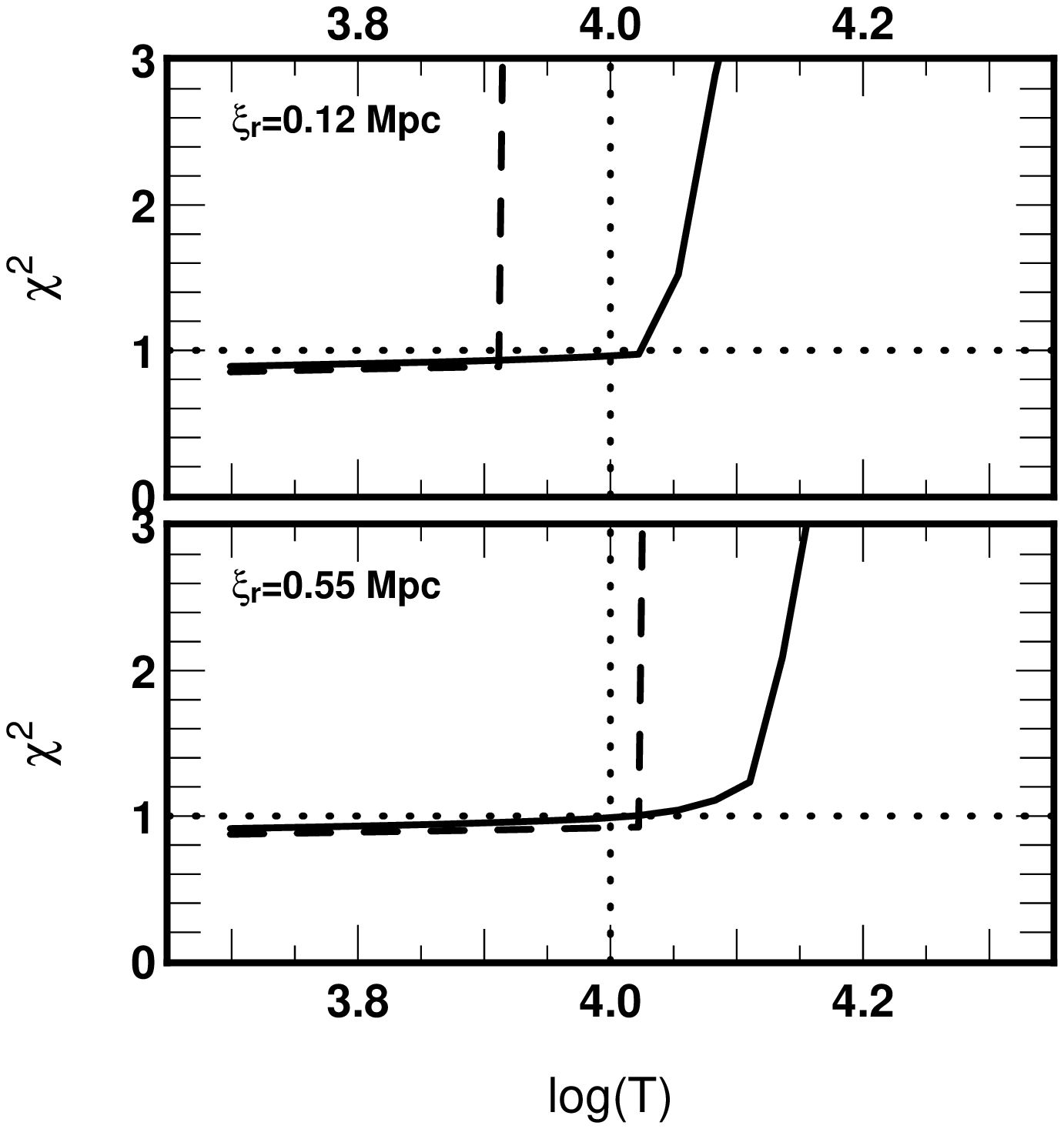}
\caption{Result of the fit  of a synthetic spectrum  
constructed from a simple 1D density field (Fig A.2)
 and assuming a temperature-density relation defined by \Eq{eos} with $\beta_0$~=~0.25 and $\overline{T}_0$~=~10\,000 K. The inversion is performed for 
the density field, for different  values of 
$\overline{T}$ while $\beta=\beta_0$ is fixed. 
The final minimum reduced $\chi^2$ is plotted 
versus $\overline{T}$.  The horizontal and vertical dotted lines 
correspond to $\chi^2=1$ and $\overline{T}=\overline{T}_0$ respectively.
The intersection of the curves with the $\chi^2$=1 line
defines the borderline. The different curves correspond
to different values of the priors as defined by the matrix $\M{C}_0$, i.e.
its variance $\sigma_p$ and correlation length, $\xi_r$ (see 
\Sec{method}).  $\xi_r=0.12-0.55$ Mpc comoving (top to bottom panel)
and $\sigma_p=0.2-0.5$ (solid and dashed lines).
In this range, the value of $\overline{T}$ on the borderline 
is constant within $10\%$. 
}
  \label{f:eos_param}
\end{figure}
\subsection{Application to synthetic spectra}
\subsubsection{Definition of a borderline in the ($\beta$,$\overline{T}$)
plane}
We first try to recover the initial
couple ($\beta_0$,$\overline{T}_0$) used to generate one of 
the synthetic spectra (see \Sec{data}). For this, we fix 
a couple ($\beta$,$\overline{T}$) and inverse the spectrum for
the density field. We derive the minimum reduced $\chi^2$ value 
obtained by fitting the model to the data,
i.e. $(\M{D}-g(\M{M}))^{\perp}.\M{C}_d^{-1}.(\M{D}-g(\M{M}))$, see \Sec{method}. 
The result of fitting the 
synthetic spectrum F$_1$ is shown in \Fig{chi_plane}. 
The ($\beta$,$\overline{T}$) plane is divided into two regions with,
respectively, $\chi^2$~$<$~1 and $\chi^2$~$>$~1, separated by a 
well-defined approximately straight borderline. 
The correct value of $\overline{T}$ lies on the $\chi^2=1$ borderline.
This is expected as explained in Appendix A.
\parn
We have investigated the dependance
of the position of this borderline with respect to
the parameters of the inversion,
namely the correlation amplitude ($\sigma_p$) and
correlation length ($\xi_r$) of the prior matrix $\M{C}_0$.
$\xi_r$ should be of the order of  the spectral resolution  
(here, the pixel corresponds to 0.05 \AA\  or 0.047 Mpc comoving) ;
$\sigma_p$ controls the fluctuations of the density
between two iterations. $\sigma_p$
 should be neither too small to allow 
for the recovery of the high density peaks, nor too large
for stability of the algorithm.
It can be seen in \Fig{eos_param} that
for $0.2\le  \sigma_p \le 0.5$ and 
$0.12\le \xi_r ({\rm Mpc\ com.}) \le 0.55$, the $\chi^2=1$ border 
corresponds to the
same $\overline{T}$
within $10\%$. We use $\sigma_p=0.25 {\rm\ and\ } \xi_r=0.2$.
\parn
Note, however,
that a high  $\chi^2$ value can 
be the consequence of only a small portion of the
spectrum beeing poorly fitted.
This implies that, when fitting real data,  the borderline can be considered as a lower 
limit for the temperature at fixed $\beta$.
\parn
It remains that  the inversion problem is degenerate: it is always possible to find 
a good fit to the data with any given value of $\beta$ provided
$\overline{T}$ is changed. We have tried to find a systematic approach to 
be able to recover from the synthetic data the temperature 
at mean density  $\overline{T}$ that was used to 
create the data.
\subsubsection{Where is the strongest constraint on $\overline{T}$ ?}
\begin{figure}
\includegraphics[width=8.5cm,angle=-0]{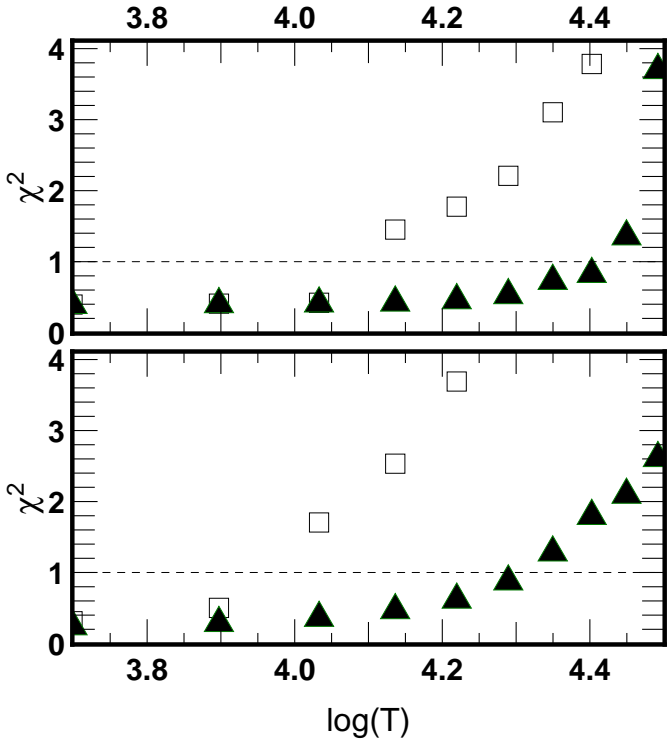}
\caption{Minimum reduced $\chi^2$ versus log~$\overline{T}$ obtained after
fitting 20~\AA~ wide portions of the synthetic 
spectrum F$_1$ (upper panel) or 
 of the HE~1128$-$1628 spectrum (lower panel). The portions are chosen so that
the minimum value of the normalized flux is smaller than 0.1 (open squares)
or is in the range 0.2$-$0.3 (filled triangles). The reconstruction is
performed for $\beta$~=~0.25. Note that in the simulation (upper panel),
$\overline{T}$~=~10000~K.
}  
\label{f:eos_structure}
\end{figure}

Zaldarriaga et al. (2000) have shown that the flux power spectrum
is affected by changes in the temperature only for overdensities between 
0.9 and 1.8. It is likely
that not all the absorption features
are well suited to constrain the temperature and one can wonder which 
features constrain the temperature best. Indeed, low density regions are 
associated with diffuse gas in voids where temperature broadening is washed 
out by the smoothness of the density field itself. This suggests that the 
strongest constraint on the temperature determination comes from 
moderately saturated lines. 
\parn
To check this, we divided the spectrum in sub-spectra, of typical width 
20~\AA. In each sub-spectrum we define the maximum optical depth,
$\tau_{\rm max}$, corresponding to a normalized flux $F_{\rm min}$. 
As argued previously, it is expected that the sub-spectra where 
$F_{\rm min}$ is large will poorly constrain the temperature. We therefore 
limit the inversion to sub-spectra where some condition on $F_{\rm min}$
is imposed. We have checked that the width  of the sub-spectra is not a 
critical parameter for the procedure. It must be large enough to include 
an important range of apparent optical depths but not too large, in order to 
keep enough weight on large optical depths.
\parn
\Fig{eos_structure} shows the results of the inversion of the sub-spectra 
of the synthetic spectrum F1, varying $\overline{T}$ while keeping $\beta$ 
equal to 0.25 (which is the correct  initial value). The reduced $\chi^2$ is 
plotted versus the mean temperature for $F_{\rm min}$~$< 0.1$ (only strong 
features are considered; results are shown as open squares in the figure) 
and for 0.2~$<$~$F_{\rm min}$~$< 0.3$ (lines of intermediate saturation only 
are considered; results are shown as filled triangles in the figure). It is 
apparent that, for weak lines, $\chi^2$ remains smaller than one for a 
very wide range of temperatures. Instead, for strong lines, $\chi^2$ 
is larger 
than one for all temperatures larger than the correct value 
$\overline{T}$~=~10$^4$~K. Note that a normalized flux of 0.2 corresponds to 
an overdensity of $\rho/{\bar \rho}\sim~2$.
 Because of this behaviour, only sub-spectra
with $F_{\rm min}$~$< 0.2$ are used in the following 
to constrain  $\overline{T}$.
\subsubsection{Estimation of the temperature $\overline{T}$}
The reduced $\chi^2$ is computed for each sub-spectrum, labelled $i$, and each couple
($\beta$,$\overline{T}$). This defines the temperature on the borderline, 
$\overline{T}_{\rm \beta,i}$, for each sub-spectrum. 
The highest $\overline{T}_{\rm \beta,i}$ correspond to
segments where 
the density field is so smoothed that the effect of temperature
is very small.
This introduces a systematic bias against low 
temperatures.

\begin{figure}
\includegraphics[width=8.5cm,angle=0]{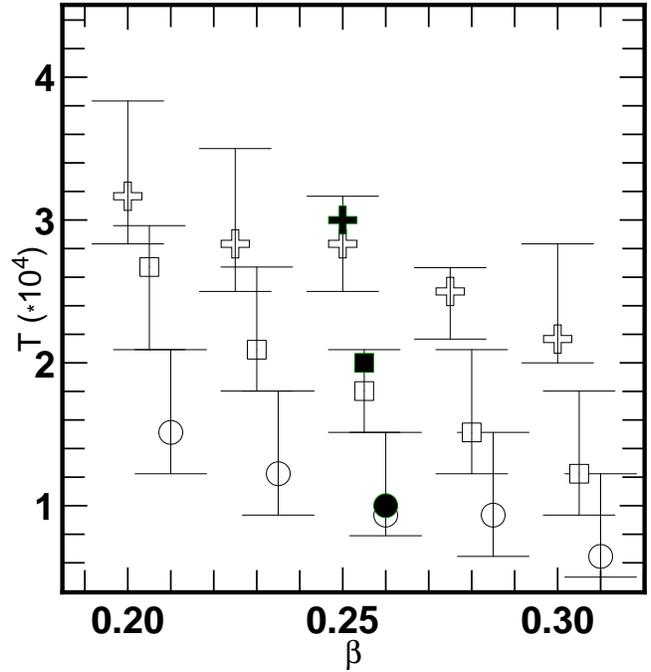}
\caption{Recovered parameters of the temperature-density relation  for different analytical spectra. 
Each of the spectra are generated using the same density field and 
$\beta$~=~0.25 but with three different temperatures: 
$\overline{T}=10\,000\ {\rm K}\ {\rm (circles),\ } 20\,000\ {\rm K}\ {\rm (squares),\ } 
30\,000 \ {\rm K}\ {\rm (crosses)}$. Results from inversions are shown as
open symbols; the initial parameters are shown as filled symbols.
See text for the definition of the estimator. The largest uncertainty in
the determination of the absolute temperature from the inversion comes
from the degeneracy between $\beta$ and $\overline{T}$.}
\label{f:simu_eos}
\end{figure}

Using the synthetic spectrum, we check that, 
for the correct 
value of $\beta$, an appropriate estimator of $\overline{T}_0$ is 
the median of  the $1^{\rm st}$ quartile  of the cumulative distribution,
$P(\overline{T})$, defined as the region
where $0.<P(\overline{T})<0.25$.
This is 
illustrated in \Fig{simu_eos} where this estimator of $\overline{T}_0$
is plotted versus $\beta$ for the three synthetic spectra F1 (10~000~K),
F2 (20~000~K) and F3 (30~000~K).
The error bars represent the range of temperatures in the $1^{\rm st}$ quartile.
The three initial
temperatures are recovered with this estimator for the correct 
value of $\beta$~=~0.25.
%
%
%
%
\subsection{Application to HE~1122$-$1628 and discussion}
\begin{figure}
\includegraphics[width=8.5cm]{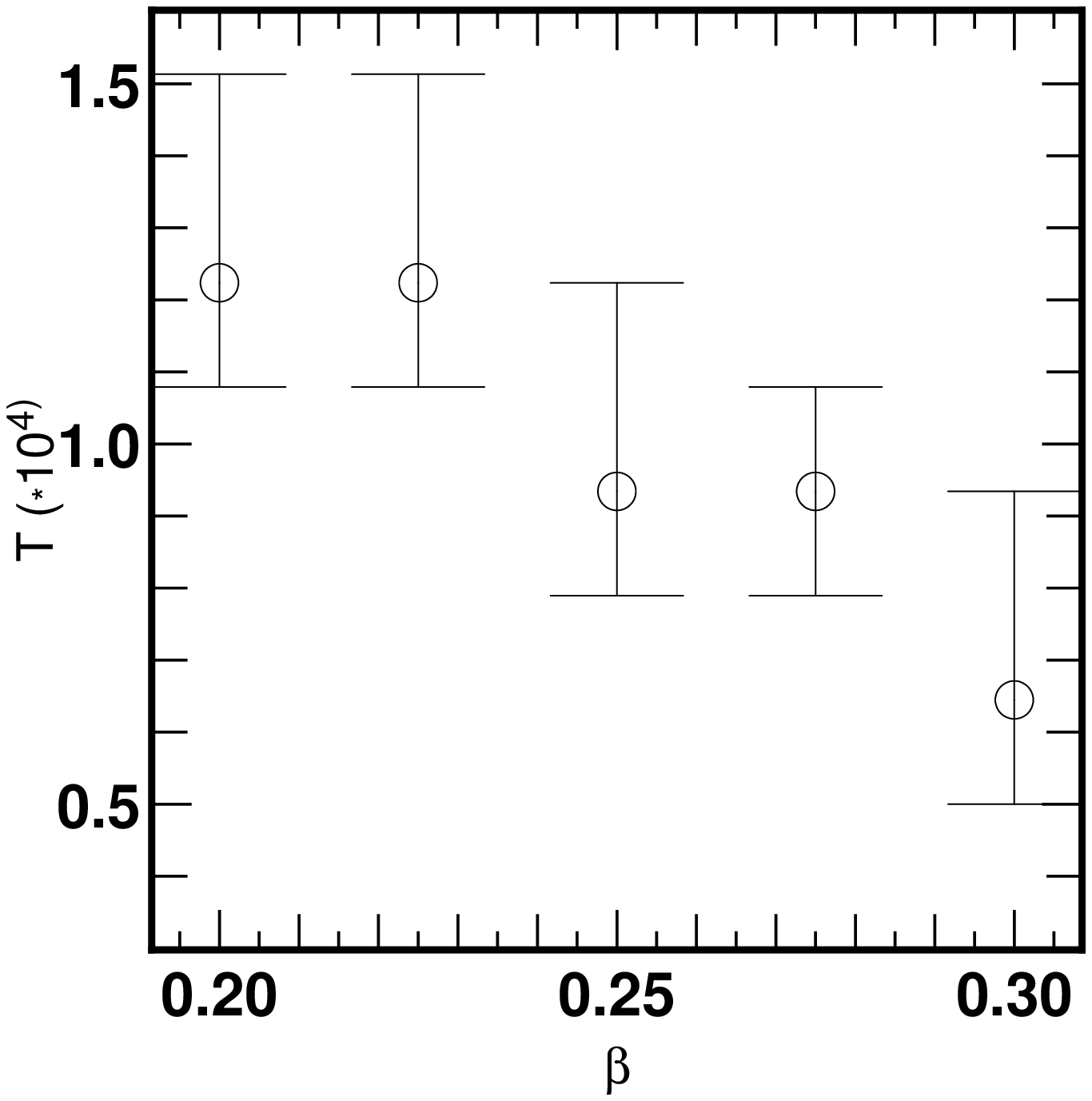}
\caption{Parameters of the temperature-density relation 
recovered by inversion of the UVES
spectrum of HE~1122-1628. The 
 temperature at mean density 
at z $\simeq 2$ is found in the range 6\,000$-$15\,000 K. 
The trend shown in this figure is close to that
found in the F1 spectrum (see \Fig{simu_eos}) where $\overline{T}$=10\,000 K.
}
\label{f:q1122_eos}
\end{figure}

The same procedure is applied to the UVES spectrum of HE~1122$-$1628.
It can be seen in \Fig{eos_structure} 
(lower panel) that, as already noted 
with analytical
data,
the fit of strong lines is more 
constraining than the fit of weak lines. 
\parn
Results on the determination  of $\overline{T}$
along the line of sight to HE~1122$-$1628 is given in \Fig{q1122_eos}.
This figure (when  compared to \Fig{simu_eos}) 
suggests  that the 
temperature of the IGM at mean density 
at redshift $z$~$\sim$~2 is only slightly 
larger than 10$^4$~K. Accounting for  the uncertainty due to 
the inherent lack of knowledge on $\beta$, we conclude that  
$\overline{T}$~$\sim$~10000$_{6000}^{15000}$~K at $z$~=~2.  
%
\parn
%
%
\parn
Other methods have been applied to determine the 
temperature-density relation  of the IGM over the 
past few years. Schaye et al. (2000), Ricotti et al. (2000) and Mc Donald 
et al. (2000) deconvolve the spectra using Voigt profile fitting. They then 
consider that the temperature can be estimated from the lower bound of the
locus of observed values of the Doppler parameter and neutral hydrogen column
density. This is reminiscent of the reasoning used by Pettini et al. (1990).
The three groups derive similar results. For an average redshift of 2.4, 
they find, respectively, $\overline{T}/10^4=0.8-2$, $\beta=0.25\pm0.07$;
$\overline{T}/10^4=0.7-5$  and $\beta=0-0.3$ and $\overline{T}/10^4=1.74\pm0.19$  and $\beta=0.26\pm0.07$ .
As emphasized by Zaldarriaga et al. (2000), these authors use a lower limit 
of the width of the lines over the whole spectrum. They therefore derive a 
lower limit on the temperature.  
Another approach that does not make use of any fitting procedure  
has been developed by Zaldarriaga et al. (2000). It relies on the 
determination of the flux power-spectrum. These authors show that there is 
no power on scales smaller than the scale corresponding to thermal broadening 
$b_{\rm T}$. They find, for the same redshift, 
$\overline{T}(\beta=0)/10^4=1.8-3.2$ and 
$\overline{T}(\beta=0.3)/10^4=1.-1.6$. All analyses show a dependance of 
$\overline{T}$ over $\beta$ similar to that shown in \Fig{q1122_eos}.
\parn 
In addition, Hui \& Gnedin (1997) conclude from the results of
hydro-simulations that $\overline{T}\ {\rm and}\ \beta$ reach an asymptotic 
value at low redshift: 
$\overline{T}\propto \left(\frac{\Omega_b\ h^2}{\sqrt{\Omega_m\ h^2}}\right)
^{1/1.7}$ and $\beta>0.15\ - 0.24\ $, depending on the exact reionization 
history. This corresponds approximately to a temperature at overdensity unity 
that decreases from about 10\,000~K at $z=4$ to 8000~K at $z=2$. Note 
however that they did not include the effects of He~{\sc ii} reionisation.
\parn 
Our findings are therefore consistent with previous published 
analysis. 
 Note that the 
remaining
discrepancies in the results obtained by the various authors could be a 
consequence of the fact that a single temperature-density relation
 is assumed for the gas.
 In particular, it is probable that spatial 
variations of the ionizing flux can have a substantial influence on the 
temperature (see e.g. Zaldarriaga 2001).
 It should be possible to investigate this in more detail using our
procedure if a large number of lines of sight is observed at the same 
resolution and S/N ratio. Such data will be available   after a few years of 
observations with UVES at the VLT. Moreover, the determination of the 
characteristic temperature of the IGM over a large redshift range could 
constrain the reionisation history. Indeed, Schaye et al. (2000) have 
tentatively found a peak of $T$~=~$10^{4.4}$~K at a redshift of 3.2 (compared 
to about $10^4$~K at redshifts 3.5 and 2) and interpret this as evidence for 
He~{\sc ii} reionisation. 
%
\section{Comparison with Voigt profile fitting}
\label{s:VPFIT}
\begin{figure}
\includegraphics[width=9cm]{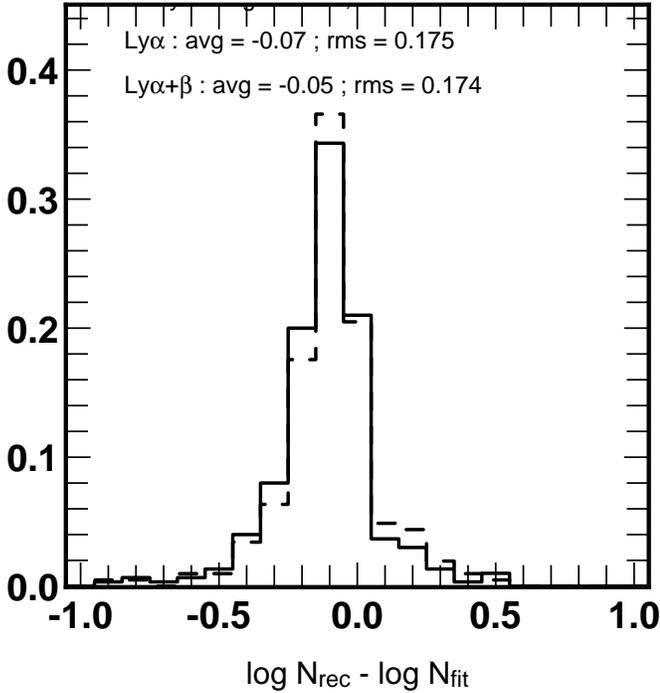}
\caption{ 
Histogram of the difference between the logarithm of the 
H~{\sc i} column densities
 obtained using the Voigt profile fitting procedure ($N_{\rm fit}$) on the one hand, and the inversion 
method ($N_{\rm rec}$) 
 on the other hand, from the fit of the Lyman-$\alpha$ forest only 
(solid line) 
or for both the Lyman-$\alpha$ and Lyman-$\beta$ forests (dashed line),
evaluated along the \los\ to 
HE~1122-1628.
For the inversion procedure, the column densities are estimated by integrating
the recovered density over twice the Doppler parameter found by the 
Voigt profile fit. 
}
\label{f:column}
\end{figure}

\begin{figure}
\includegraphics[width=8.5cm]{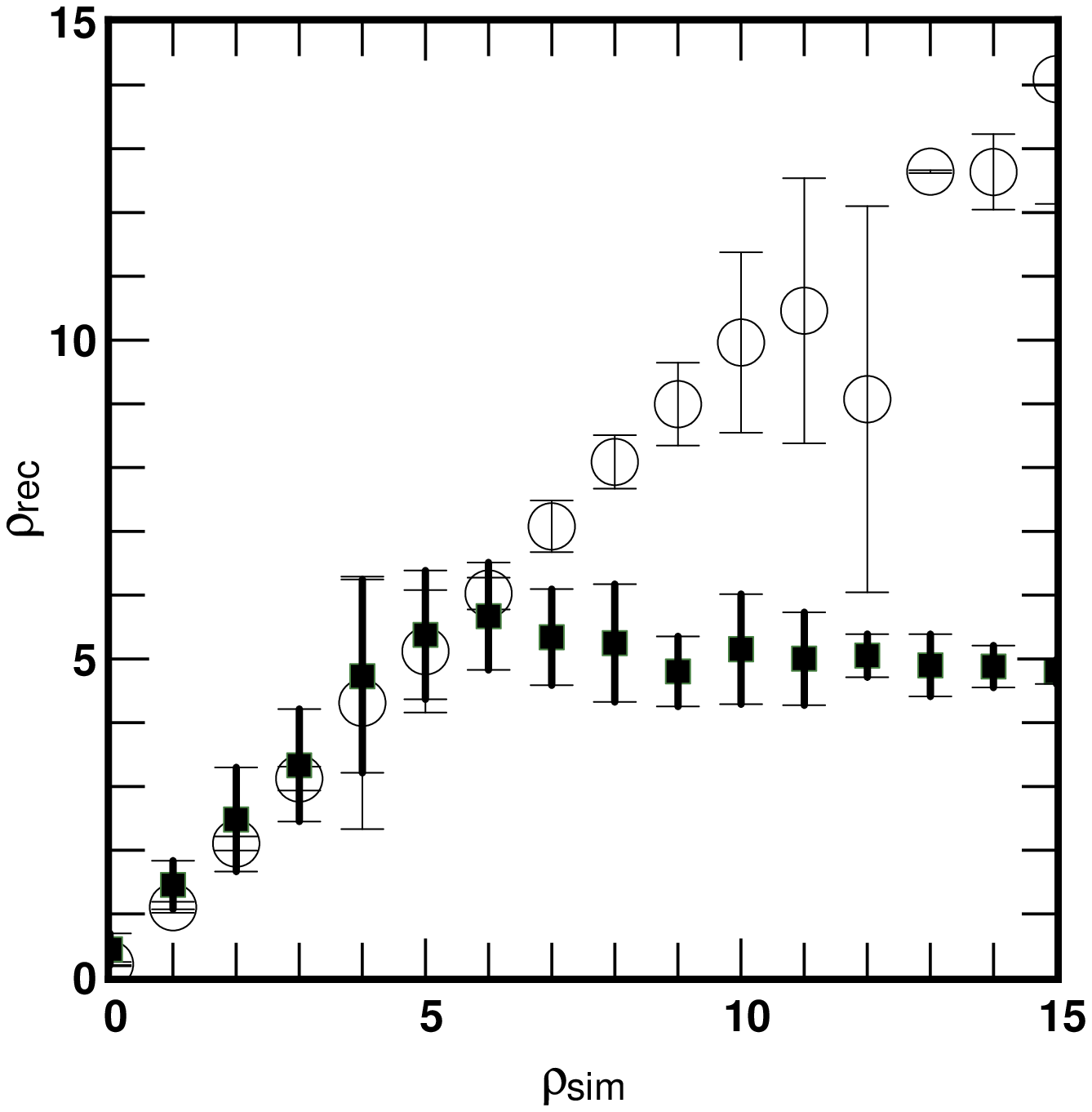}
\caption{Over-densities recovered after inversion of a synthetic spectrum
versus the over-densities used to generate the synthetic spectrum,
assuming $\overline{T}=10^4 {\rm K}, \beta=0.25$.} 
When  only Lyman-$\alpha$ data is available for  the inversion procedure (filled squares), 
overdensities larger than 5 cannot be recovered because of saturation
effects. This situation is very much improved when using the Lyman-$\beta$
forest data (open circles).
\label{f:density}
\end{figure}

\begin{figure}
\includegraphics[width=8.5cm]{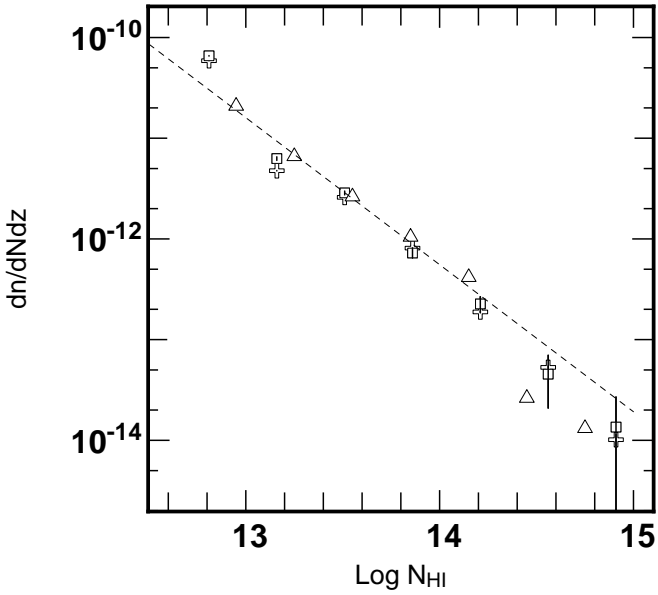}
\caption{
Column density distribution at z\ $\simeq$ 2.3, 
corrected for incompleteness, 
as measured toward HE~1122-1628 from the inversion procedure (crosses)
and the Voigt profile procedure (squares). 
Also plotted is the distribution obtained by Kim et al. (1997, triangles).}
\label{f:cdd}
\end{figure}

\begin{figure}
\includegraphics[width=9cm]{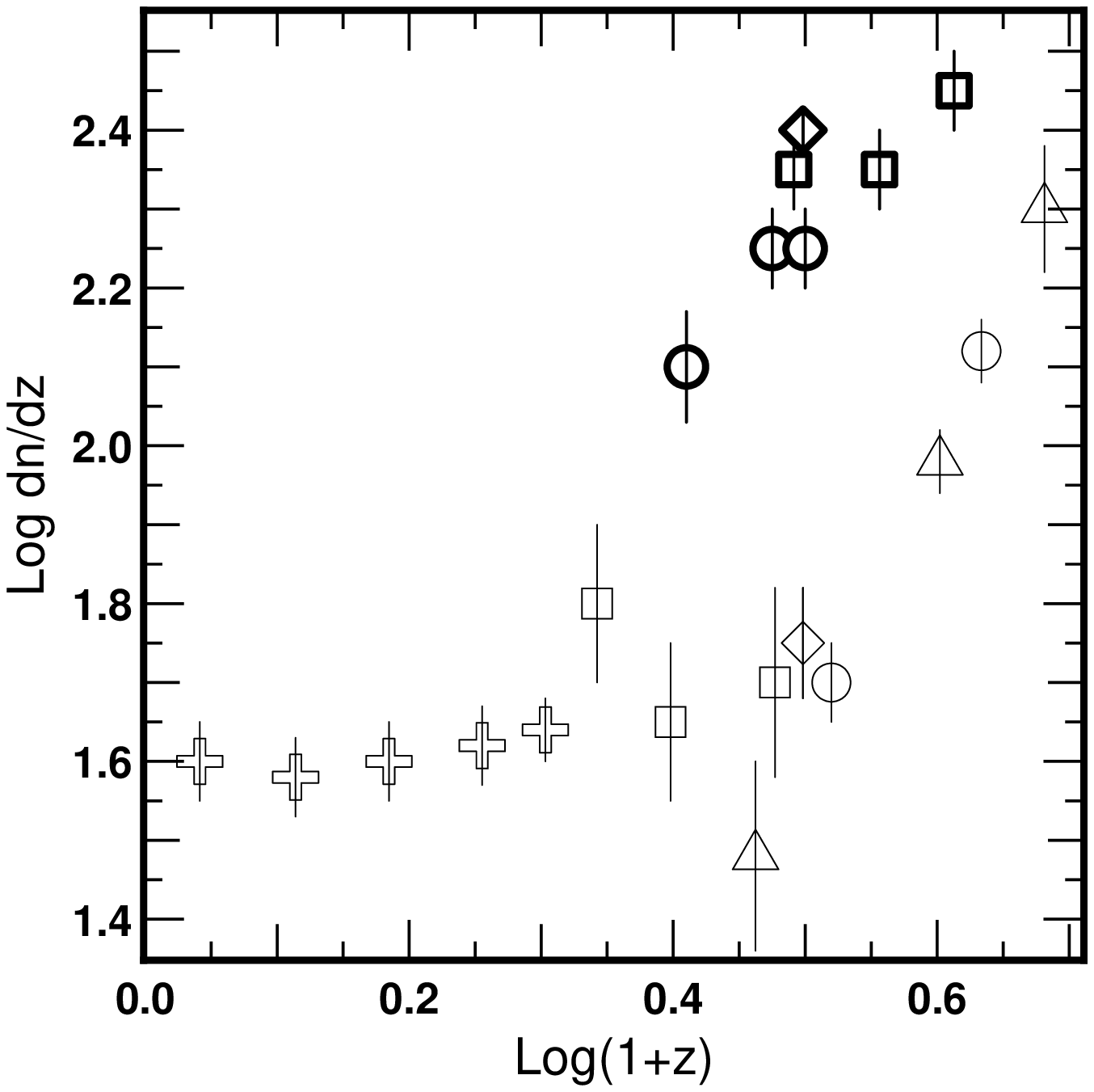}
\caption{
Number density of lines versus redshift for different column density 
thresholds: log $N$(H~{\sc i}) $>$ 13 (Kim et al. 1997, thick squares; Kim 
et al. 2000, thick circles; this work, thick losange), 
log $N$(H~{\sc i}) $>$ 14 (Savaglio et al. 1999, squares); Weymann et al.
1998, crosses; Kim et al. 1997, circles; Giallongo et al. 1996, triangles;
this work, losange).
}
   \label{f:evolution}
\end{figure}

In this Section, we compare the column densities evaluated along the line of 
sight to HE~1122$-$1628 from the fit of the Lyman-$\alpha$ forest using the 
inversion method on the one hand and VPFIT, the automatic Voigt profile 
procedure (Carswell et al. 1987) on the other hand. The latter procedure 
decomposes the spectrum in discrete absorption 
components, whose  Doppler 
parameters and column densities  are fitted to the data. Our method 
derives the density in each velocity pixel and we must integrate this density
over some velocity range, chosen somewhat arbitrarily, to recover a column 
density  and compare it to  VPFIT's. At the position of each cloud 
defined by VPFIT, we integrate over twice the Doppler parameter of the cloud. 
Table~A.1 summarizes the results.
\parn 
\Fig{column} shows the histograms of the values taken by the ratio between 
the two column densities when only the Lyman-$\alpha$ line is considered 
(solid line) and when the Lyman-$\beta$ line is used in addition to constrain
the fit (dashed line). The mean difference between the two estimations is
about 15\%, well within the uncertainties due to the measurements themselves.
The inversion method usually underestimates the high column densities. This 
is a consequence of the lack of information in the saturated part of the 
corresponding absorption lines. In addition, the fact that the procedure 
reconstructs a continuous field minimizes the importance of the wings of the 
line. The situation is improved, as expected, when the Lyman$-\beta$ line is 
used to constrain the fit. The mean difference is then reduced to about 10\%.
\parn
The improvement in the inversion of the density field when including the 
Lyman-$\beta$ forest is emphasized in \Fig{density}. We inverse one of the 
synthetic spectra for which we know the real overdensity in each 
pixel. \Fig{density} gives the reconstructed overdensity versus the 
overdensity used to simulate the spectrum. When the Lyman-$\alpha$ line only 
is used, the density field is recovered up to overdensities slightly larger 
than 5, whereas when the Lyman-$\beta$ forest is used in addition, 
overdensities up to 10 are recovered.
\parn
\Fig{cdd} shows the H~{\sc i} column density distribution derived from 
the inversion of the Lyman-$\alpha$ and Lyman-$\beta$ forests observed toward 
HE~1122$-$1628 together with the results from the automatic fitting of the 
Lyman-$\alpha$ forest using VPFIT and the data points from Kim et al. (1997). 
A correction has been applied at the low column density end to take into 
account incompleteness induced by the blending of weak lines 
with strong ones (Kim et al. 1997).
There is a remarkable agreement 
between the inversion method, VPFIT and previous observations.
\parn 
The number of absorption lines per unit redshift is given in \Fig{evolution} 
as a function of redshift. Although \ previous analyses have 
concentrated on strong lines, log~$N$(H~{\sc i})~$>$~14, most of the 
constraints 
{on the IGM} come from weaker lines. Our results at 
$z$~$\sim$~2 for log~$N$(H~{\sc i})~$>$~13 and log~$N$(H~{\sc i})~$>$~14,
 together with other observational points 
(Giallongo et 
al. 1996, Kim et al. 1997, Weymann et al. 1998, Savaglio et al. 1999, 
 Kim et al. 2000), nicely fall on top of the predictions by 
Riediger et al. (1998).
\parn
%

%
\parn 
In this Section, we have not yet  taken advantage of the true capabilities of 
the Bayesian method. Indeed, we first inverted the spectrum to a density field 
and then reprojected the result into discrete clouds.
 In the following, we concentrate on  other questions,
taking full advantage of  the new method.
%
%
\section{Structures in the strong lines}
 \label{s:structure}
As long as only unsaturated lines are considered, the 
corresponding structures are clearly
identified in the spectrum. The same structures are lost, however,
 when blended 
with saturated lines {whose width  is } often of the order of 
100~km~s$^{-1}$ (see below). The nature of the strong lines 
(log~$N$(H~{\sc i})~$>$~14) is unclear and they could be associated with 
filaments, halos with rotation velocities less than about 100~km~s$^{-1}$ 
and/or external regions of galactic halos (see e.g. M\"ucket et al. 1996). 
Observations of adjacent lines of sight have concluded that these lines arise 
in complexes with transverse dimensions larger than 200~kpc. To gain insight 
into the small-scale structure of these complexes, we can try to invert the 
corresponding Lyman-$\alpha$ lines, adding information from all observed lines 
in the  Lyman series and when available from the associated C~{\sc iv} lines.
The inversion is performed in redshift space.
\subsection{Inversion of strong lines}
We have selected for further analysis the 12 lines seen in the spectrum of 
HE~1122$-$1628 which have log~$N$(H~{\sc i})~$>$~14 and a large enough 
redshift ($z_{\rm abs}$~$>$~2.03) so that the Lyman-$\beta$ line is redshifted
in the observed wavelength range. 
A more accurate Voigt 
profile fit has been performed on these individual lines, including whenever 
possible other lines in the Lyman series (mostly Lyman-$\beta$ and 
Lyman-$\gamma$). 
Also, we have inverted the flux with the 
Bayesian method including all pieces of information available. 
Since the 
associated C~{\sc iv} lines are most of the time too weak, we do not include
the C~{\sc iv} profile as a constraint for the inversion and keep it instead 
for an a-posteriori discussion of the C~{\sc iv}/H~{\sc i} ratio.
\parn
%
\begin{figure}
\includegraphics[width=8.5cm]{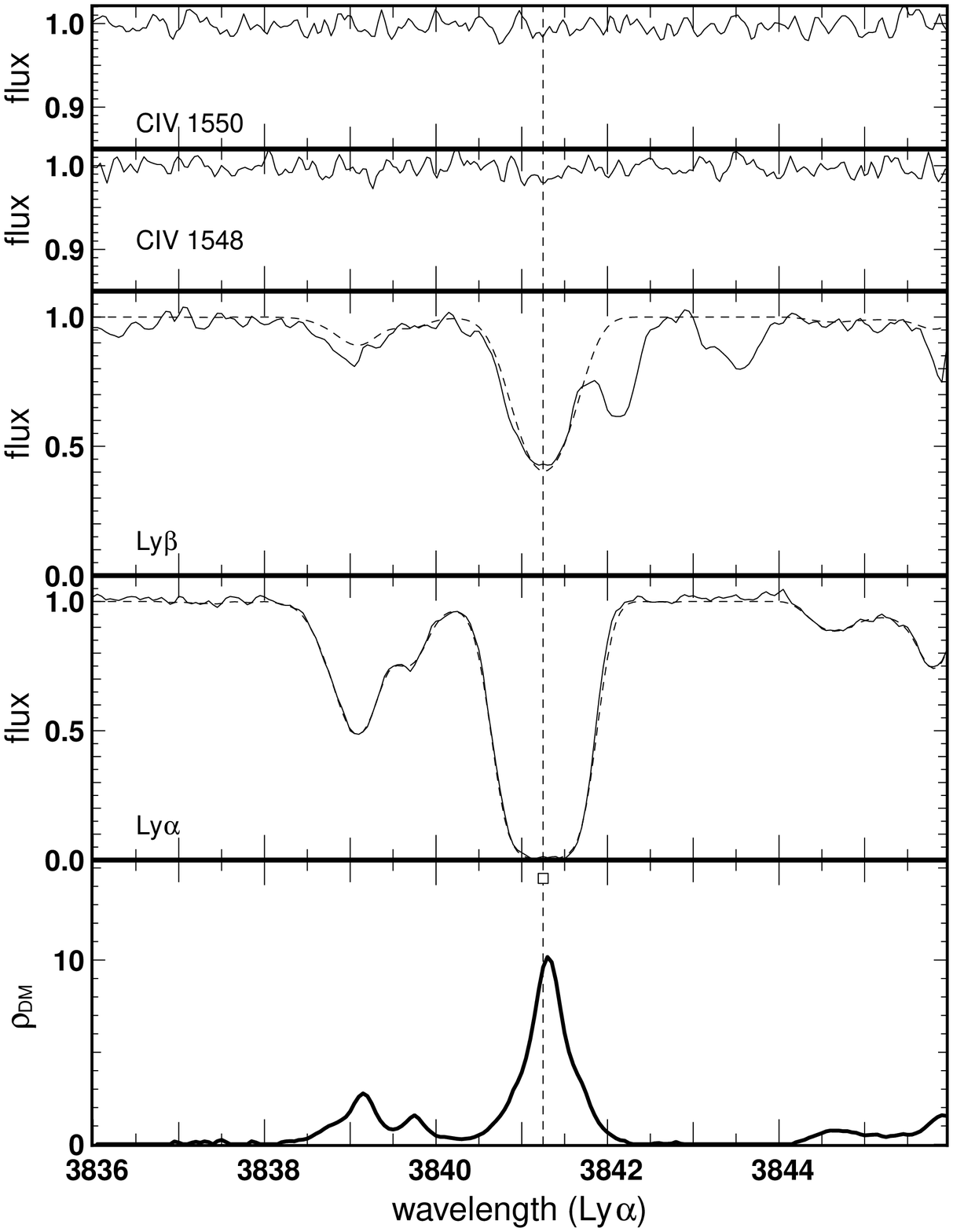}
\caption{Example of the inversion of a saturated line, using both the 
Lyman-$\alpha$ and Lyman-$\beta$ profiles. The temperature-density relation
 used during inversion 
is defined by $\beta=0.25$ and $\overline{T}=8\,000$ K (\Eq{eos}).
The recovered density is shown in the bottom panel. 
The other panels shows from bottom to top the Lyman-$\alpha$,
Lyman-$\beta$, C~{\sc iv}$\lambda$1548 and C~{\sc iv}$\lambda$1550 profiles.
The reconstructed flux after inversion is overplotted as dashed lines.
Note that the blending of the Lyman-$\beta$ forest 
is ignored by the inversion.
Vertical lines show the position of components identified by the Voigt profile
fitting procedure.
}
\label{f:structure_lyb_3841}
\end{figure}

\begin{figure}
\includegraphics[width=8.5cm]{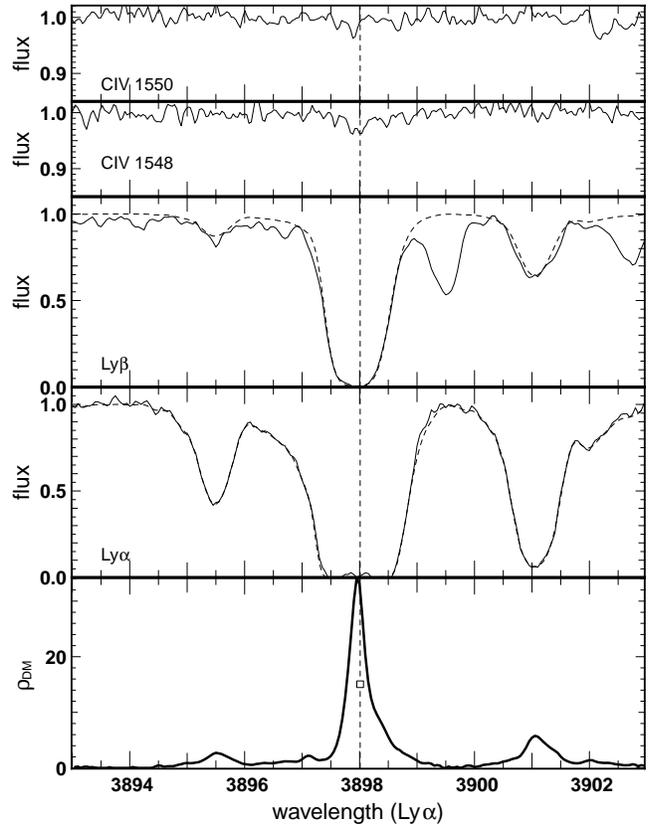}
\caption{Same as \Fig{structure_lyb_3841} for another system.
 Here $\overline{T}=11000$ K.
Note that the density is reconstructed even where the
Lyman-$\alpha$ forest is saturated, thanks to the constraints
from the Lyman-$\beta$ forest.}
\label{f:structure_lyb_3898}
\end{figure}

\begin{figure}
\includegraphics[width=8.5cm]{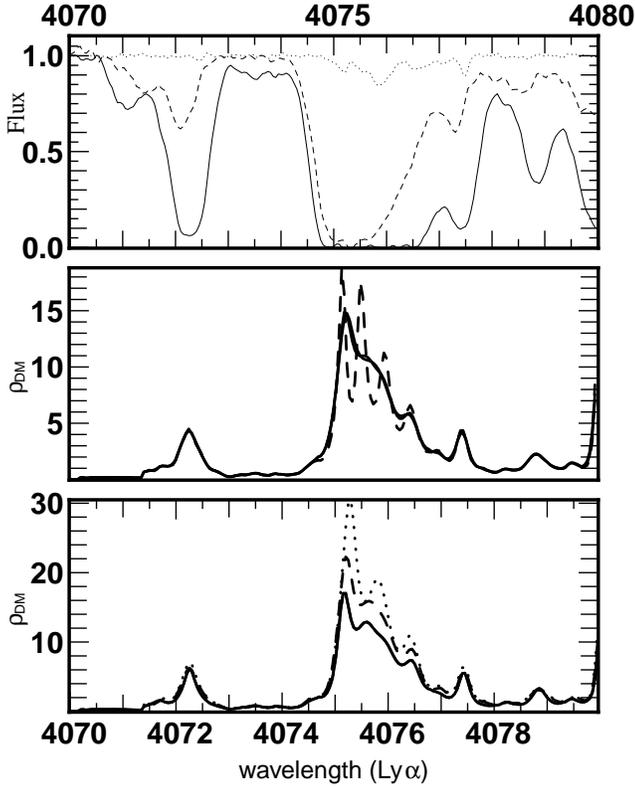}
\caption{Illustration of the influence of the a-priori correlation length
(see text) and the temperature-density relation
 on the recovered density in the 
Lyman-$\alpha$ complex at $\lambda$~$\sim$4076~\AA~ (see also 
\Fig{structure_lyb_4076}). {\sl Top panel}: Lyman-$\alpha$, Lyman-$\beta$ 
and C~{\sc iv} absorption profiles are shown as, respectively,  solid, 
dashed and dotted lines. {\sl Middle panel}: recovered density for different 
correlation lengths, 0.05 \AA\ (dashed line) and 0.1-0.2 \AA\ (solid line, 
both profiles are undistinguishable). Note that the pixel size is 0.05 \AA. 
{\sl Bottom panel}: recovered density for three different values of the
temperature-density relation parameters; the three ($\beta$,$\overline{T}$) couples are 
on the borderline defined in \Fig{q1122_eos}: solid line, 
$\beta=0.2,\  \overline{T}=15\,000$ K, dashed line, 
$\beta=0.25,\  \overline{T}=10\,000$ K, dotted line, 
$\beta=0.3,\  \overline{T}=5\,000$ K.}
\label{f:structure_eos_delta}
\end{figure}
%
\begin{figure}
\includegraphics[width=8.5cm]{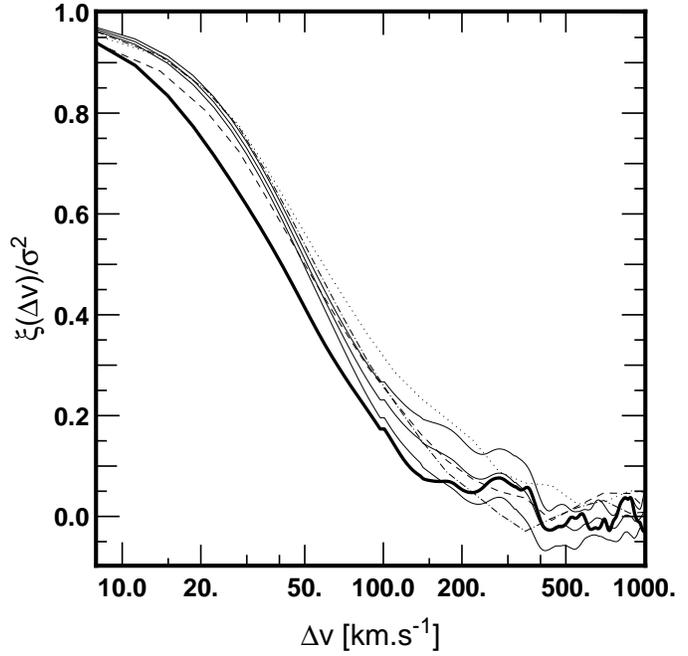}
\caption{Two-point correlation functions of the flux in HE~1122$-$1628 
at $z$~$\sim$~2. (light solid lines); the three lines correspond to the 
function itself and the function $\pm$1$\sigma$. 
The dashed lines correspond to the measurements by McDonald et al. (2000)
at $z$~=~3.89, 3.0  and 2.41 are shown as, respectively 
  dashed, dotted and dash-dotted lines.
The thick solid line is  the function calculated for the recovered density.
}
\label{f:correlation}
\end{figure}
%
%
\parn
In principle, the Lyman-$\beta$ line can always be blended with an intervening
Lyman-$\alpha$ line at lower redshift. We therefore define 
\begin{equation}
\tau_{Ly\beta}={\rm min}(\tau_{z\beta},\tau_{z\alpha}*f_\alpha\lambda_\alpha/f_\beta\lambda_\beta),
 \EQN{tau_beta}
\end{equation}
where $\tau_{z\beta}$ and $\tau_{z\alpha}$ are the observed optical depth at 
the position of the Lyman-$\beta$ and Lyman-$\alpha$ absorptions respectively;
$f_{\alpha}$, $f_{\beta}$, $\lambda_\alpha$ and $\lambda_\beta$ are the 
oscillator strengths and rest-wavelengths of the Lyman-$\alpha$ and 
Lyman-$\beta$ lines. We then invert 
 $\tau_{Ly\beta}\ 
{\rm and\ } \tau_{Ly\alpha}$ simultaneously (with the same relative  weight 
for the two lines). 
\parn
%
\begin{figure}
\includegraphics[width=8.5cm]{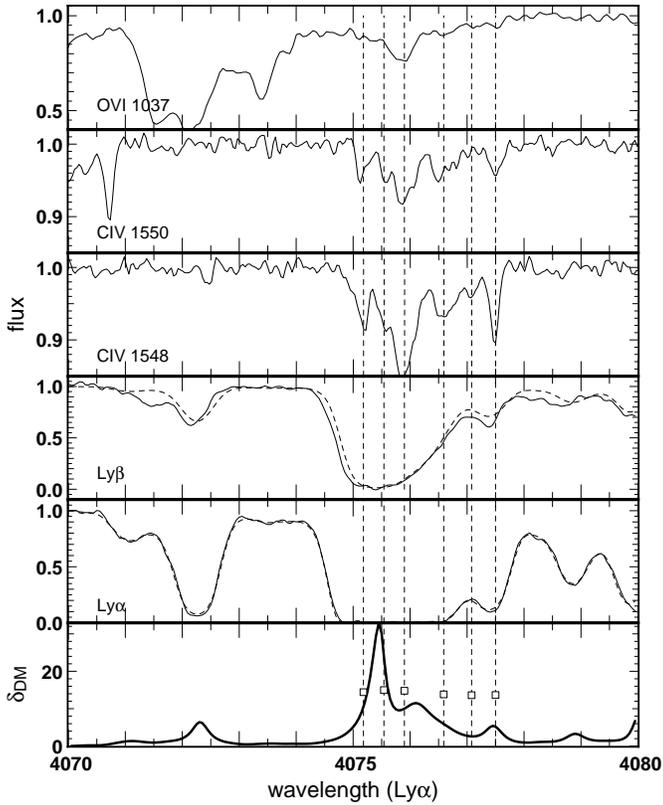}
\caption{Same as \Fig{structure_lyb_3841} for the complex at 
$\lambda$4076 \AA. 
The C~{\sc iv} is strong enough to allow us to study  the ratio
C~{\sc iv}/H~{\sc i} (\Fig{CIV_HI_4076}). Note the shift between the H~{\sc i} and C~{\sc iv}
strongest peaks (the position of the density peaks does not 
depend strongly on the parameters of the inversion, 
see \Fig{structure_eos_delta}). Note also the presence of
C~{\sc iv} in the wings of H~{\sc i} profile (at $\lambda \sim$ 4076.5 \AA, 
marker B in \Fig{CIV_HI_4076}).
The O~{\sc vi} profile is also shown in the upper panel. The similarity of
the O~{\sc vi} and C~{\sc iv} profiles 
is a signature of predominance of photo-ionization 
in the ionization balance.}
\label{f:structure_lyb_4076}
\end{figure}
\begin{figure}
\includegraphics[width=8.5cm]{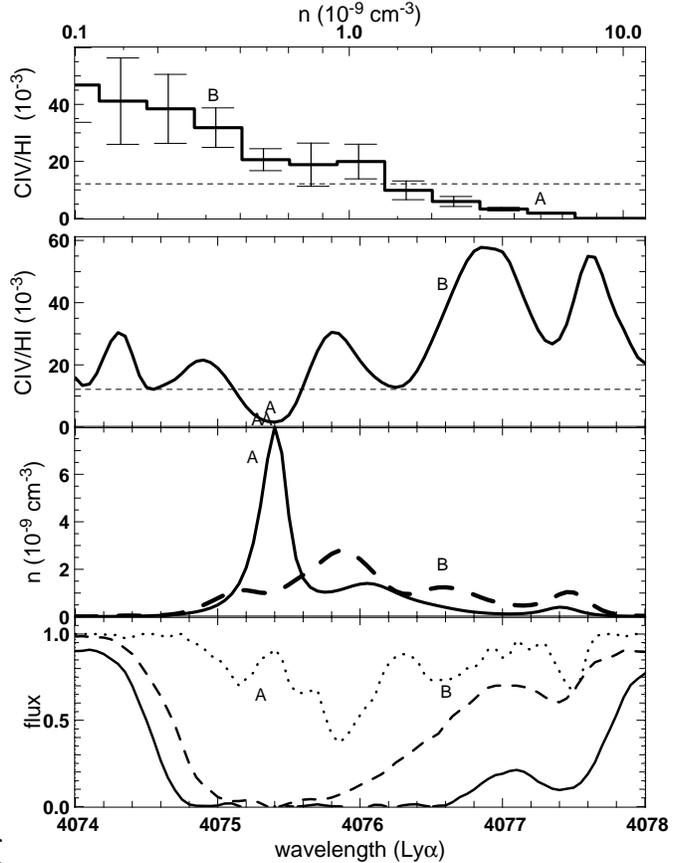}
\caption{Determination of the C~{\sc iv}/H~{\sc i} ratio 
along the absorption profile of the strong complex at $\lambda$4076 \AA\ (see
also \Fig{structure_lyb_4076}). {\sl Bottom panel}: 
Lyman-$\alpha$ (solid line), Lyman-$\beta$ (dashed line) and C~{\sc iv} 
(scaled by a factor 4, dotted line) absorption profiles. 
{\sl Second panel}: recovered  H~{\sc i} (solid line) and C~{\sc iv} 
(scaled by the average C~{\sc iv}/H~{\sc i} ratio, dashed line) density 
(see text for details about the reconstruction). {\sl Third panel}:
C~{\sc iv}/H~{\sc i} ratio; the horizontal line indicates the ratio
of the C~{\sc iv} and H~{\sc i} integrated densities. {\sl Top panel}:
the C~{\sc iv}/H~{\sc i} ratio is plotted versus the H~{\sc i} density. 
The markers A and B correspond to positions discussed in the text.
}
\label{f:CIV_HI_4076}
\end{figure}
%
\begin{figure}
\includegraphics[width=8.5cm]{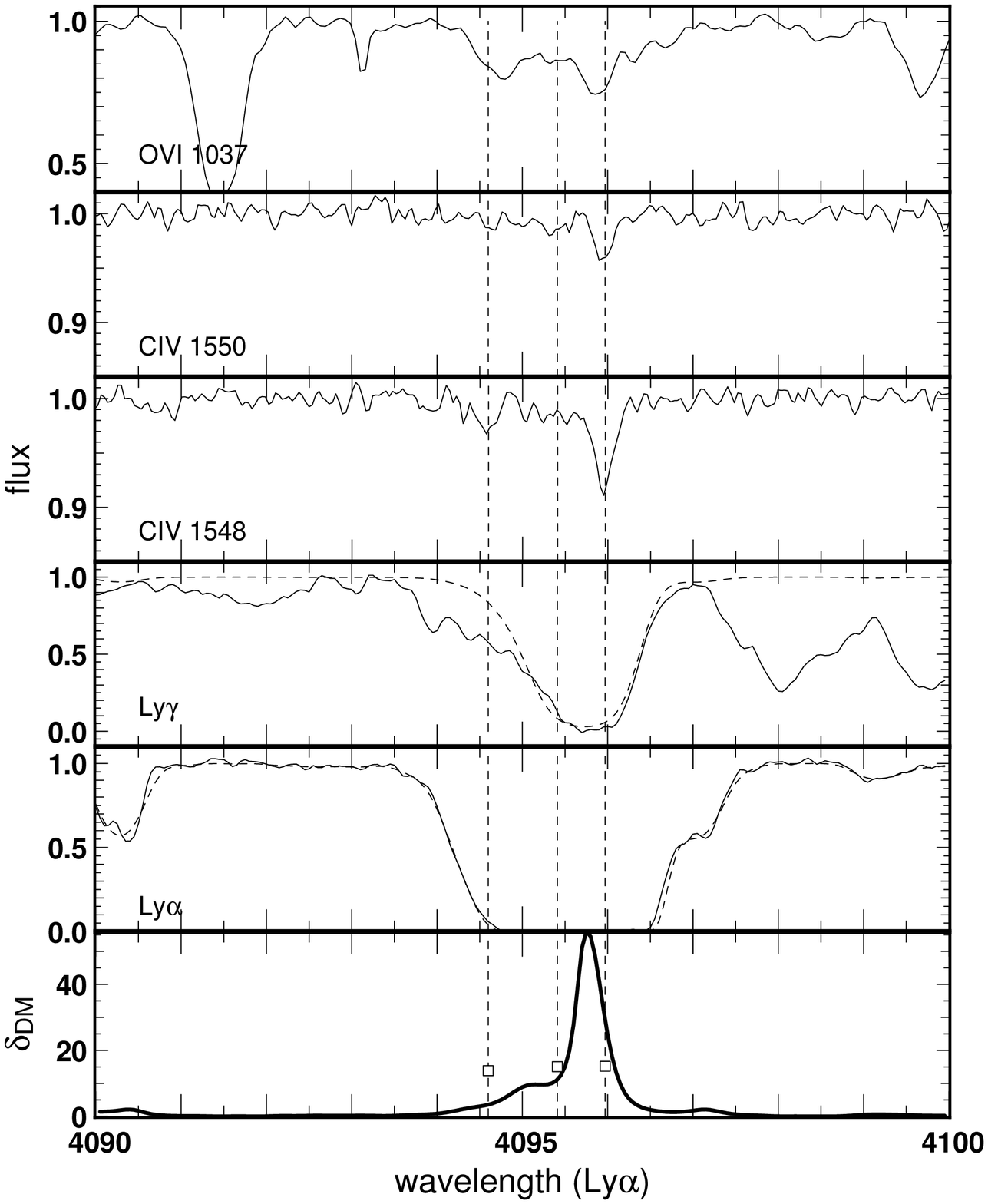}
\caption{Same as \Fig{structure_lyb_4076} for the complex at
$\lambda$4096~\AA. 
}
\label{f:structure_lyc_4096}
\end{figure}

\begin{figure}
\includegraphics[width=8.5cm]{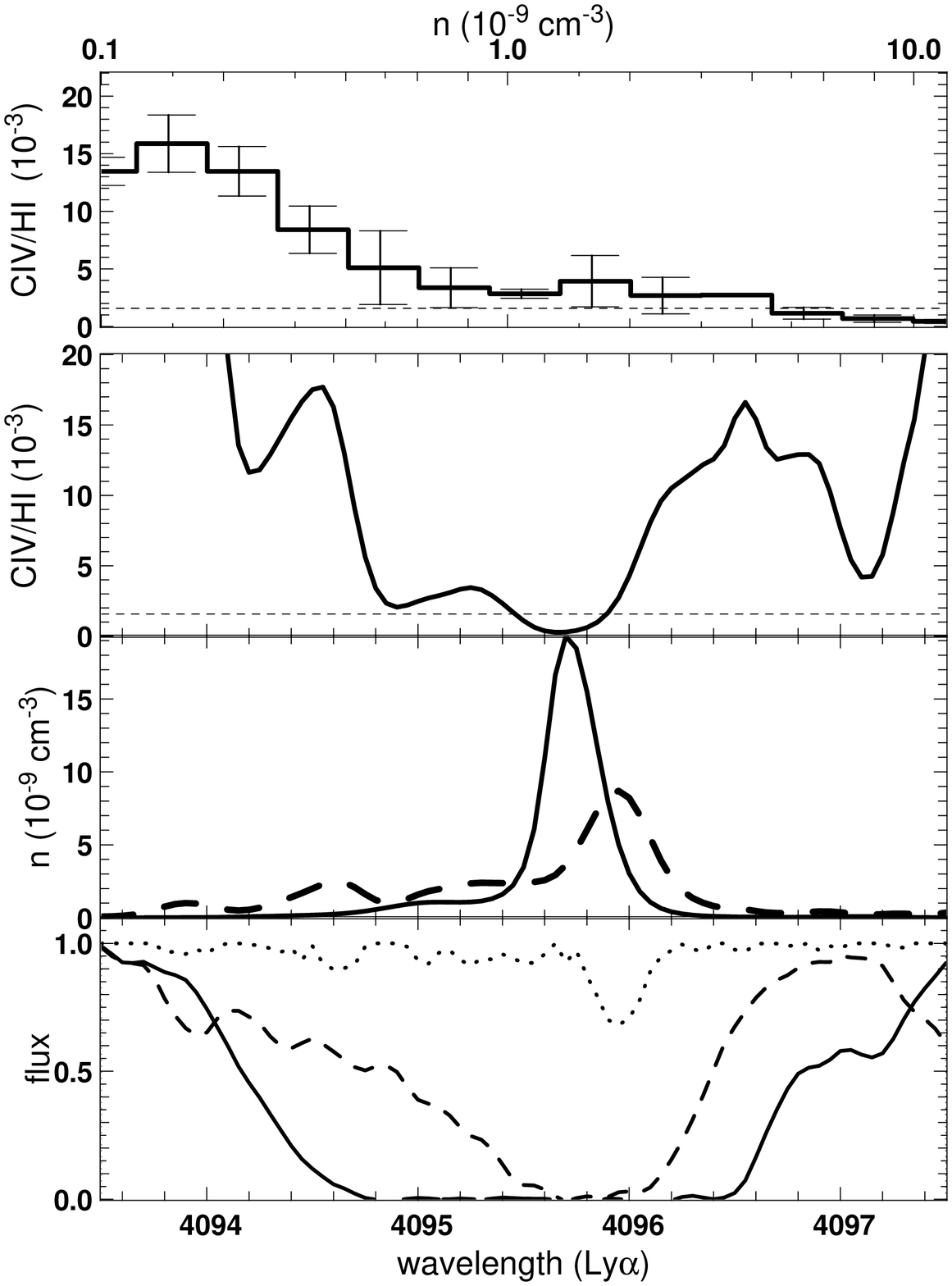}
\caption{Same as \Fig{CIV_HI_4076}
 for the complex at $\lambda$4096~\AA\ 
 (see also \Fig{structure_lyc_4096}). 
}
\label{f:CIV_HI_4096}
\end{figure}
%
\parn
A few examples of the fits and reconstructions are given in 
Figs~\ref{f:structure_lyb_3841},\ref{f:structure_lyb_3898},\ref{f:structure_lyb_4076},\ref{f:structure_lyc_4096}. 
From bottom to top, the different panels correspond to (i) the reconstructed 
density, (ii) the Lyman-$\alpha$ absorption profile, 
(iii) the Lyman-$\beta$ or the Lyman-$\gamma$ profiles, 
(iv) the associated C~{\sc iv}$\lambda\lambda$1548,1550 doublet
and (v) when possible, the associated O~{\sc vi} absorption. The 
fit models are overplotted on the observed Lyman profiles as 
dashed lines. 
The positions of the components used in the Voigt profile fits are marked
by vertical lines. 
\parn
\subsection{Validation}
The inversion procedure uses a correlation length to propagate the information
along the line of sight. It is important to check the robustness of the 
results when changing this parameter. For this, we invert the Lyman-$\alpha$ 
structure at 4075 \AA \ different correlation lengths (see 
\Fig{structure_eos_delta}, middle panel). We use the Lyman-$\beta$ and 
Lyman-$\alpha$ profiles to constrain the fit. It can be seen that for 
correlation lengths larger than the resolution of the spectrum ($\sim$0.1~\AA~
at this wavelength), the results are weakly dependent on the exact value of 
the parameter. It is only for very small correlation lengths (the pixel 
size is 0.05 \AA), that the structures become artificially peaked. In the 
following, we always use a correlation length of 0.2 \AA.
\parn
As the temperature-density relation is determined for the whole spectrum, it is possible
that the global solution is not representative of the physical state of the
gas in one particular place. To estimate the influence of a possible variation
of the temperature-density relation
 from one place to the other, we have inverted the 
same Lyman-$\alpha$ feature at 4075~\AA~ using three 
different values of the couple
($\beta$,$\overline{T}$), found on the borderline defined in \Sec{eos}.
Note that the values used for the three inversions are extreme and correspond
to variations of a factor of three in $\beta$ and $\overline{T}$. It can be 
seen in \Fig{structure_eos_delta} (bottom panel) that although the absolute 
amplitude of the density field
varies by a factor of $\sim$2 from one inversion to the other, the 
shapes of the density profiles are very similar and all the structures are 
recovered. 
\subsection{Correlation function}

The two-point correlation function is defined as, 
\begin{equation}
\xi(\Delta v)=<\delta F(v)\delta F(v+\Delta v)>.
\end{equation} 
The function computed for the density field reconstructed from the inversion
of the HE~1122$-$1628 spectrum 
is plotted in \Fig{correlation} together with the correlation function
of the flux.
The curves are normalized to 
$\xi(0)=1$. To compute errors, we used the  
bootstrap method  presented in, e.g.,  McDonald et al. (1999)
and found that 
the correlation 
function is not corrupted by the amount of noise.
On small scales, 
the shape of the function is probably related to the 
internal structure of the Lyman-$\alpha$ clouds (or more realistically
Lyman-$\alpha$ complexes), the dimension of which
is expected from $N$-body simulations to be of the order of 1~Mpc
($\simeq~130\ {\rm km\ s}^{-1}$).
Observations of adjacent lines of sight also indicate that the absorptions are
correlated on scales of the order of 0.2 to 1~Mpc (e.g. Petitjean et al. 1998, 
D'Odorico et al. 1998, Dinshaw et al. 1998, Crotts \& Fang 1998). However, it 
must be noted that the radius derived from analysis of the metal lines 
associated with the Lyman-$\alpha$ forest are much smaller (Hu et al. 1995; 
see next Section).
\subsection{The C~{\sc iv}/H~{\sc i} ratio}
The determination of the amount of metals present in the Lyman-$\alpha$ forest 
attracts much interest because this is a key issue for understanding how the 
first objects in the Universe were formed. Since the first observations of 
metals (mostly C~{\sc iv}) in at least half of the Lyman-$\alpha$ lines with 
log~$N$(H~{\sc i})~$>$~14.3 (Cowie et al. 1995, Songaila \& Cowie 1996), 
efforts have been concentrated towards constraining the metal content of the
gas with smaller column densities. As direct detection of the corresponding 
very weak C~{\sc iv} lines is difficult even if the physical conditions are 
the same at different column densities, a statistical approach has been used. 
Ellison et al. (1999, 2000) conclude that observations are consistent with a 
constant C~{\sc iv}/H~{\sc i}~$\sim$~$2 \times 10^{-3}$ ratio down to very 
small column densities. Deriving a carbon metallicity from this is quite 
uncertain, however. Using hydro-simulations including photo-ionization, 
Rauch et al. (1997) claim that at $z$~=~3, the column density ratios of the 
different ionic species can be well reproduced if a mean metallicity of 
[Z/H]~=~$-$2.5 is assumed. 
\parn
Out of the 12 clouds with $N$(H~{\sc i})~$>$~10$^{14}$~cm$^{-2}$ seen toward
HE~1122$-$1628, 7 (60\%) have associated C~{\sc iv} absorption.
The average C~{\sc iv}/H~{\sc i} ratio computed after Voigt profile fitting
of the absorptions is $4\times 10^{-3}$. The optical depth ratio 
$\tau_{\rm CIV}$/$\tau_{\rm HI}$, where $\tau_{\rm CIV}\ {\rm and\ \tau_{HI}}$
are defined as in \Eq{tau_beta} for all available components is approcimately
 constant 
and corresponds to C~{\sc iv}/H~{\sc i}$\simeq 2.5\ 10^{-3}$
 for $\tau_{\rm HI}>1$ and slightly smaller for 
$\tau_{\rm HI}<1$. 
\parn
It has already been noted by Songaila \& Cowie (1996) that the 
C~{\sc iv}/H~{\sc i} ratio is not constant within a strong complex and seems 
to be larger in the wings of the strong lines. Using the inversion method we 
are in a position to study the variations of the C~{\sc iv}/H~{\sc i} ratio 
within the strong complexes. For this we must invert, not only
 the Lyman series as 
described above but also the C~{\sc iv} absorption.
Unfortunately, due to the faintness of the C~{\sc iv} absorptions,
 the required S/N 
ratio 
is very large.
Although the S/N ratio of the  HE~1122$-$1628 spectrum is very good, 
it is high enough only for the $z_{\rm abs}$~=~2.35 and $z_{\rm abs}$~=~2.37 
systems (see Figs.~\ref{f:CIV_HI_4076},\ref{f:CIV_HI_4096}). We invert the 
absorptions for the H~{\sc i} density using Lyman-$\alpha$,$\beta$ for the 
4076 \AA\ complex and Lyman-$\alpha$,$\beta$,$\gamma$ for the 4096 \AA\ 
complex. The C~{\sc iv} density is obtained from the 
C~{\sc iv}$\lambda\lambda$1548,1550 doublet.
The H~{\sc i} particle density is of the order of 10$^{-9}$~cm$^{-3}$
which is what is expected from simulations (typically
$n_{\rm H}$~$\sim$~10$^{-4}$~cm$^{-3}$ and H~{\sc i}/H~$\sim$~10$^{-5}$,
see e.g. Bi \& Davidsen 1997, Hellsten et al. 1997). It must be noted however 
that this number is obtained assuming that the density is constant over the 
pixel size, $\simeq$~12~kpc at $z$~=~2. The C~{\sc iv}/H~{\sc i} ratio varies from 
about 0.001 to 0.01 within the complexes. Finally, it can be seen that there 
are slight velocity shifts ($\sim$10~km~s$^{-1}$) between the peaks in the 
C~{\sc iv} and H~{\sc i} density profiles. 
\parn
The most interesting observation, however, is that there is some hint of
an anti-correlation between C~{\sc iv}/H~{\sc i} and $n_{\rm HI}$ (see 
Figs.~\ref{f:CIV_HI_4076},\ref{f:CIV_HI_4096} upper panel where 
 the C~{\sc iv}/H~{\sc i} ratio is plotted versus the 
H~{\sc i} density). It is difficult to assess the
anti-correlation as the S/N ratio in the C~{\sc iv} profile is probably not 
high enough and more important, the statistics are  poor. However, it is 
apparent on the profiles that (i) the strongest H~{\sc i} components have 
C~{\sc iv}/H~{\sc i}~$\sim$~a few 10$^{-3}$ (e.g. region A) 
and (ii) there is gas in the wing 
of the profiles of lower H~{\sc i} optical depth and much larger 
C~{\sc iv}/H~{\sc i} ratio (e.g. region B).
 A straightforward interpretation would be that 
the carbon metallicity is not homogeneous in the cloud. A more simple 
explanation however is that this is a consequence of temperature fluctuations.
Indeed C~{\sc iv}/H~{\sc i} is quite sensitive to the temperature for 
$T$~$\sim$~10$^{5}$~K (see e.g. Rauch et al. 1997). Because of high 
temperature dielectronic recombinations, C~{\sc iv}/H~{\sc i} increases with 
temperature. If the anti-correlation is real, this implies that the temperature of 
the gas increases for decreasing H~{\sc i} particle density {\sl and} is close
to 10$^{5}$~K.
\parn
Interestingly enough, it can be noted that for collisional ionization, 
C~{\sc iv}/C has a maximum $\sim$0.35 at $T$~$\sim$~10$^{5}$~K 
(Arnaud \& Rothenflug 1986) and C~{\sc iv}/H~{\sc i}~$\sim$~2$\times$10$^{-2}$
for [C/H]~=~10$^{-2.5}$. However, if collisional ionization dominates, no 
O~{\sc vi} is expected in the gas, as the dominant ionization stage of oxygen 
is O~{\sc iv} at this temperature. On the contrary, in the gas we observe, 
it can be seen in Figs.~\ref{f:CIV_HI_4076},\ref{f:CIV_HI_4096} that 
O~{\sc vi} is 
indeed detected and {\sl has a similar apparent optical depth as} C~{\sc iv}.
Note that we plot in each case the line of the O~{\sc vi} doublet which is
not blended. Moreover, as far as the S/N ratio allows the comparison, the 
C~{\sc iv} and O~{\sc vi} profiles are strongly correlated. 
{\sl This is a clear signature 
of the predominance of photo-ionization in the ionization balance of the gas}.
%
%
\section{Conclusion}
\label{s:conclusion}
%
We have applied the Bayesian inversion procedure developped by
Pichon et al. (2001) to the high resolution and high S/N ratio spectrum
of HE~1122$-$1628 obtained  during Science Verification of UVES
at the VLT and made
publicly available by ESO. Assuming a temperature-density relation
$T=\overline{T} \left( \frac{\rho_{\rm DM}(\M{x})}
{\bar{\rho}_{\rm DM}}\right)^{2\beta}$ for the gas, we invert the 
spectrum for the density field $\rho(x)$ and 
recover a grid  of $\chi^2$ versus the two parameters
$\beta$ and $\overline{T}$. 
\parn
We certify the method by showing that the inversion procedure applied to
synthetic data enables us to infer the correct temperature when $\beta$ is 
given the value used to generate the data. There is an intrinsic degeneracy
however between $\beta$ and $\overline{T}$ which is the source of most of
the uncertainty in the determination of the temperature of the intergalactic
medium. We find $\overline{T}$~$\sim$~10000$_{6000}^{15000}$~K at $z \sim
2$.
\parn
The results of fitting the Lyman-$\alpha$ forest with the new method or
with the automatic Voigt profile fitting procedure VPFIT (Carswell et al. 
1987), when expressed in terms of column densities, are found to be very 
similar. The new procedure however has the advantage of reconstructing the
continuous density field. The latter can be recovered up to 
$\rho / {\bar \rho}$~$\sim$~5 when
the Lyman-$\alpha$ forest only is used. Above this value, saturation prevents
any reliable determination of the density. Overdensities up to 10 can be
recovered however when the Lyman-$\beta$ absorption is used.
\parn
We investigate the structure of the complexes with log~$N$(H~{\sc i})~$>$~14
by inverting the Lyman-$\alpha$ and Lyman-$\beta$ lines and therefore
derive the two-point correlation function of the dark matter on scales
smaller than 1~Mpc. Out of the twelve lines with log~$N$(H~{\sc i})~$>$~14,
seven have detected associated C~{\sc iv} absorption. 
The  ratio C~{\sc iv}/H~{\sc i}, derived from 
the optical depth ratio $\tau_{\rm CIV}$/$\tau_{\rm HI}$, is roughly 
constant, $\simeq 2.5\ 10^{-3}$ for $\tau_{\rm HI}>1$ and slightly smaller for 
$\tau_{\rm HI}<1$. In two of the complexes, we are able to inverse the
C~{\sc iv} absorption independently. This reveals the presence of fluctuations
as large as one order of magnitude in the C~{\sc iv}/H~{\sc i} ratio.
This, together with the presence of O~{\sc vi} with the same absorption
profile as C~{\sc iv} indicates that the gas is photo-ionized and 
at a temperature close to 10$^5$~K. 
\parn
Throughout this work, we did not take into account redshift distorsion.
It is expected to modify quantitatively the result of \Sec{eos}
since the lines get thinner or broader depending of the
exact distribution of the peculiar velocities. This must be investigated 
in detail and is beyond the scope of this paper. However, the conclusion of 
\Sec{structure} about the relation between C~{\sc iv} and
H~{\sc i} should be valid even in real space since velocities 
modify both profiles in the same way.   
\acknowledgements{We thank C\'edric Ledoux for providing us with the
normalized spectrum of HE~1122$-$1628, Bastien Aracil for the fit of the 
Lyman-$\alpha$ forest using VPFIT the package provided by Bob Carswell;
and R. Srianand and T.R. Choudhury for the analytical spectra. We acknowledge 
the efforts of the UVES team to finalize such a beautiful instrument and thank 
the Science Verification team, J. Bergeron, S. Cristiani, S. D'Odorico, 
T. Kim, for the data used in this paper.
}
%
%


\appendix

\def\Mfunction#1{{\rm #1}}
\def\Mvariable#1{{\rm #1}}
\def\Muserfunction#1{{\rm #1}}
\def\multsp{}
\def\ExponentialE{e}
\def\ScriptCapitalA{{\cal A}}      
\def \DifferentialD {{\rm d}}

\section{Formal thermal broadening deconvolution}
For a polytropic temperature-density relation, $ T={{\overline T}{{\rho (x)}^{-2 \beta }}}$,
the optical depth,$ \tau(w)$, is related to the underlying density via
\Eq{eqfun},
where we set $A({\overline z})$ to 1 for simplicity. 
Let us explore the r\'egime where \Eq{eqfun} is non singular ($\beta \neq 0
$). We argue here that, in  this r\'egime, 
  at a fixed $\beta$,  the inversion
problem  sketched in  \Sec{eos}  is well  paused:  i.e. there  is a  unique
solution for the underlying density field, given the measured $\tau$.
%
\subsection{Asymptotic Expansion; Gaussian profile}
%
Let us take the Taylor series of 
the integrant of \Eq{eqfun} with respect to $\beta$
\begin{eqnarray}
 && {
\frac{1}{{\sqrt{\overline      T}}}}{{\rho       (x)}^{\alpha      -\beta      }}
\exp\Big(-\frac{{{(x-w)}^2} }{\overline  T}{{\rho (x)}^{-2 \beta  }}\Big) \approx \nonumber \\
&&
\frac{{{\rho                   (x)}^{\alpha                  }}}{{\sqrt{\overline
T}^{3}}}\exp\Big(-\frac{{{(w-x)}^2}}{\overline T}\Big) \left[{\overline T}
-\beta  {(T-2 {{(w-x)}^2})
\log(\rho  (x)) }\right] \nonumber \\  \EQN{taylor}
\end{eqnarray}

Let us also assume for now that  the  underlying density field 
is Gaussian with an unknown  rms, $\sigma_{\delta}^2$:%
\begin{equation}
{\rho (x)=\frac{1}{{\sqrt{2 \pi }} \sigma }\exp\Big(-\frac{{x^2}}{2 
{{\sigma_{\delta}
    }^2}}\Big)}  \, . \EQN{defprof}
\end{equation} 
The optical depth associated with this density profile, \Eq{defprof} reads :
\begin{equation}
{\tau (w)=\frac{{{(2  \pi )}^{\frac{1-\alpha }{2}}}  {{\sigma_{\delta} }^{1-\alpha }}
 {\exp}  (-\alpha  {w^2}  )}{{\sqrt{2  {{\sigma_{\delta} }^2}+\alpha  {\overline  T}}}  (2
 {{\sigma_{\delta} }^2}+\alpha {\overline T})}} \Big(\frac{P(w) \beta {\overline T}}{2 {{\big(2
 {{\sigma_{\delta} }^2}+\alpha {\overline T}{}\big)}^4}}+1\Big)\,, \EQN{taup}
\end{equation}
 where

\begin{eqnarray}
P(w)&=&{{ {\overline T}   }^3}  (\log(2 \pi  ) \alpha  +1) {{\alpha
}^3}-\nonumber \\  & & 2  {{  {\overline  T}   }^2}  \big({w^2}+\big({w^2}  \alpha   -3
{{\sigma_{\delta} }^2}\big) \log(2  \pi )\big) {{\alpha }^3}-\nonumber \\  & & 
{\overline T} {{\sigma_{\delta}
}^2} \big({w^2} \alpha  (2 \log(\pi ) \alpha +\log(4)  \alpha -4)-\nonumber \\  & & 3 {{\sigma_{\delta}
}^2} (\alpha  \log(2 \pi )-1)\big) \alpha  +\nonumber \\ & &
 2 {{\big(2 {{\sigma_{\delta}  }^2}+
{\overline  T} \alpha  \big)}^2} \big(-2  \alpha {w^2}+2  {{\sigma_{\delta}  }^2}+{\overline T}
\alpha  \big) \log(\sigma_{\delta}  ) \alpha  - \nonumber \\ & & 8 {{\sigma_{\delta}  }^2} \big({{\alpha  }^2}
{w^4}+\alpha  {{\sigma_{\delta}  }^2} (\alpha  \log(2  \pi  )-5) {w^2}+\nonumber \\ & & 
{{\sigma_{\delta}  }^4}
(2-\alpha \log(2 \pi ))\big) \, . \EQN{defP}
\end{eqnarray}
When $\beta\equiv  0$, the first term  in the bracket  of \Eq{taup} vanishes;
the effect  of the temperature  and the natural  broadening of the  line are
degenerate, {i.e.}   all combinations of $\sigma_{\delta}  $ and ${  {\overline T}}$ which
leave ${2 {{\sigma_{\delta} }^2}+\alpha {\overline  T}}$ invariant give the same
optical depth profile.  \parn
Thanks  to \Eq{defP},  when  $\beta \ne  0$ the  shape  of the  line is  non
Gaussian,  with a  departure from  a Gaussian distribution 
 which  is a  function  of the
underlying temperature of the  medium.  This distortion is characteristic of
the temperature-density relation, and can in principle be disentangled.
 
To  demonstrate this  formally, we  may for  instance compute  the variance,
${{\sigma }_2}$,{and} {the} {kurtosis} ${{\sigma }_4}$, {of} $\tau (w)$ {as}
a {function} {of} ${\overline T}$ {and} $\sigma_{\delta} $ (for fixed $\alpha$ and $\beta$)
\begin{eqnarray}
{{\sigma }_2} &\equiv&  \frac{1}{\int \tau (w) \DifferentialD  w}\int \tau (w)
{w^2}\DifferentialD     w \nonumber \\  &=&\frac{{{\sigma_{\delta}     }^2}}{\alpha     }+{\overline     T}
\Big(\frac{1}{2}-\log(\sigma_{\delta} )  \beta -\frac{\beta }{2  \alpha }-\frac{1}{2}
\log(2 \pi ) \beta \Big) \EQN{defs2}
\end{eqnarray}
\begin{eqnarray}
     {{\sigma  }_4} &\equiv& 
 \frac{1}{\int \tau (w)  \DifferentialD w}\int
\tau  (w)  {w^4}\DifferentialD   w  \nonumber \\  &=&\frac{3  {{\sigma_{\delta}  }^4}}{{{\alpha
}^2}}+\nonumber \\  &&{{ {\overline T}   }^2} \Big(\frac{3}{4}-3 \log(\sigma_{\delta} ) \beta
-\frac{3  \beta  }{2  \alpha  }-\frac{3}{2} \log(\pi  )  \beta  -\frac{3}{4}
\log(4)  \beta \Big)+  \nonumber \\  && {\overline  T}  \Big(\frac{3 {{\sigma_{\delta}  }^2}}{\alpha
}-\frac{9  \beta {{\sigma_{\delta} }^2}}{{{\alpha  }^2}}-\frac{6 \beta  \log(\sigma_{\delta} )
{{\sigma_{\delta}  }^2}}{\alpha  }-\frac{3  \beta  \log(\pi )  {{\sigma_{\delta}  }^2}}{\alpha
}-\nonumber \\  &&
\ \ \ \ \ \ \ \ \frac{3 \beta \log(16) {{\sigma_{\delta} }^2}}{4 \alpha }\Big) \, . \EQN{defs4}
\end{eqnarray}

Solving for ${\overline T}$ in \Eq{defs2} and eliminating ${\overline T}$ subsequently
in  \Eq{defs4}  we  are  left  with  an implicit equation  for  $  \sigma_{\delta}  [\sigma_2,
\sigma_4,\beta ,\alpha ] $:
\begin{eqnarray}
 && \big(6 \beta (4 \alpha  \log(\sigma_{\delta} ) (\log(\sigma_{\delta} ) \alpha +\log(2 \pi
) \alpha +3) \beta +5 \beta + \nonumber \\  &&\alpha (\beta \log(2 \pi ) (\log(2 \pi ) \alpha
+6)-4)) {{\sigma_{\delta} }^4}+ \nonumber \\  && 2 \alpha \big(-6 \beta \big(4 \alpha
\log(\sigma_{\delta} ) (\log(\sigma_{\delta} ) \alpha +\log(2  \pi ) \alpha +2) \beta +3 \beta
+ \nonumber \\&& \alpha \big(\beta \big(\big({{\log}^2}(2)+\log(\pi ) \log(4
\pi  )\big)  \alpha +4  \log(\pi  )\nonumber \\ &&
\vspace{2cm} +\log(16)\big)-2\big)\big) {{\sigma  }_2}
{{\sigma_{\delta} }^2}+ \nonumber \\ && 3 {{\alpha }^2} (4 \alpha \log(\sigma_{\delta} ) \beta
+2 \beta  +\alpha (2 \log(\pi )  \beta +\log(4) \beta  -1)) \sigma _{2}^{2}+
\nonumber \\ &&  \alpha {{(2 \beta \log(\sigma_{\delta} ) \alpha  +\beta \log(2 \pi )
\alpha -\alpha +\beta )}^2} {{\sigma }_4}\big)\big)=0 \, . \EQN{ss2}
\end{eqnarray}
As expected, \Eq{ss2} becomes  singular when  $\beta  =0$ i.e.  it 
becomes:
$
{ {{\sigma }_4}=3  \sigma_{2}^{2}} \EQN{singular}
$
(this just  states that $\tau$ is also Gaussian and as such its  reduced  kurtosis vanishes)
and does  not involve $\sigma_{\delta} $ any longer.
On the other hand, when $\beta  \neq 0$, \Eq{ss2} has a solution for $\sigma_{\delta}
$ and  therefore for the corresponding  ${\overline T}$ via  \Eq{defs2}.  We have
therefore demonstrated that it is possible to disentangle natural broadening
and thermal broadening for non-singular temperature-density relations; in the
limit of small $\beta$ but non-zero $\beta$. 

The above argument obviously relies on the Taylor expansion, 
\Eq{taylor}; we have checked that carrying the expansion to 2nd order 
did not qualitatively change our conclusions.

\subsection{Finite signal to noise}
 
\begin{figure}
 \centering
\includegraphics[width=6.cm,angle=-90]{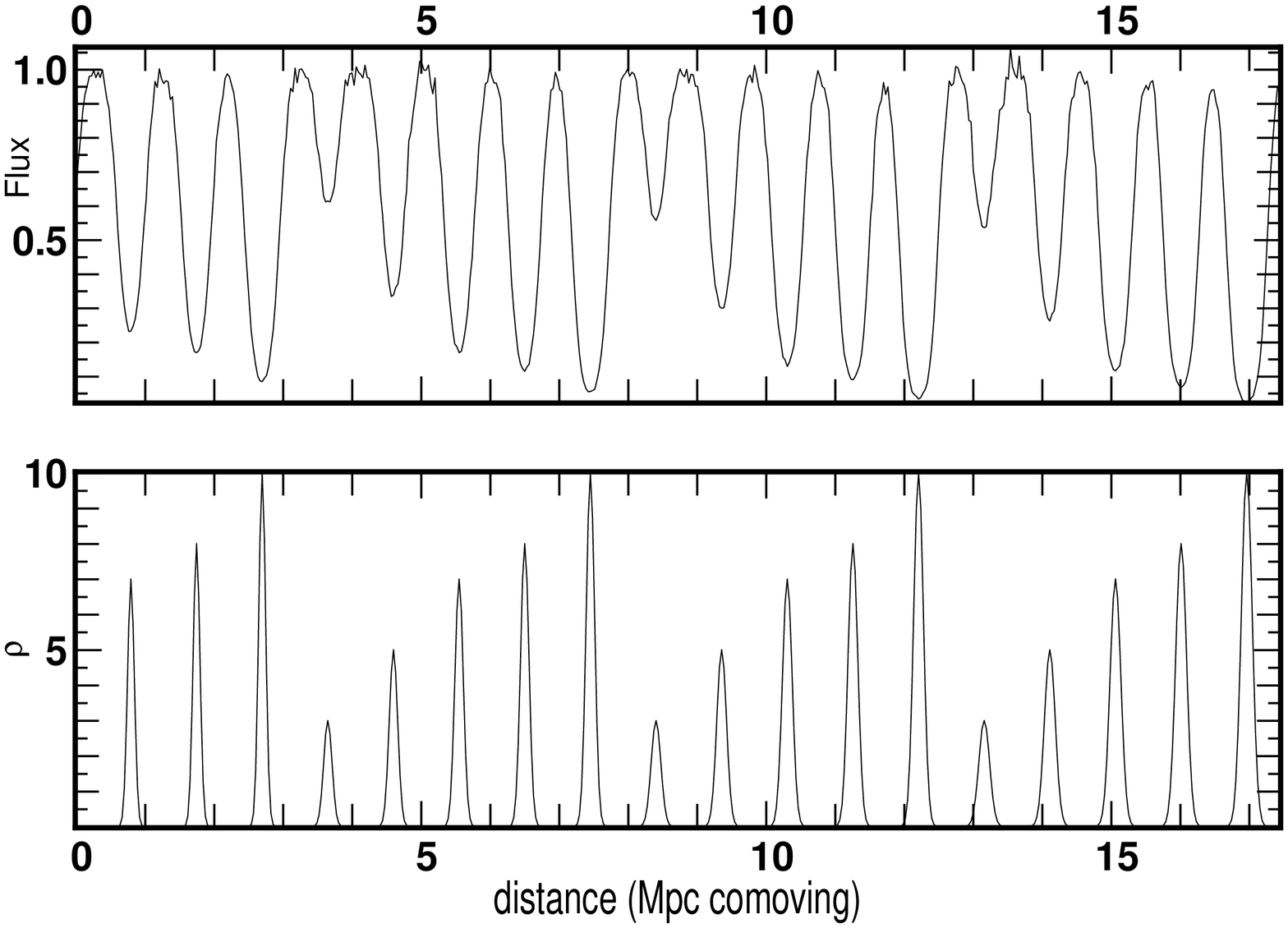}
\caption{\los\ built with a wide range of density profiles (lower panel)
and the corresponding spectrum computed with a temperature at
mean density, $\overline{T}$, of 15\,000 K and $\beta$=0.25 (upper panel).
Ten spectra are computed with a range of $\overline{T}$ 
from 5\,000 K to 30\,000 K. $\overline{T}$ is then recovered with the 
method described in \Sec{eos} (see \Fig{grid_eos_result}).
}
\label{f:grid_eos_data}
\end{figure}

\begin{figure}
\centering
\includegraphics[width=8.5cm]{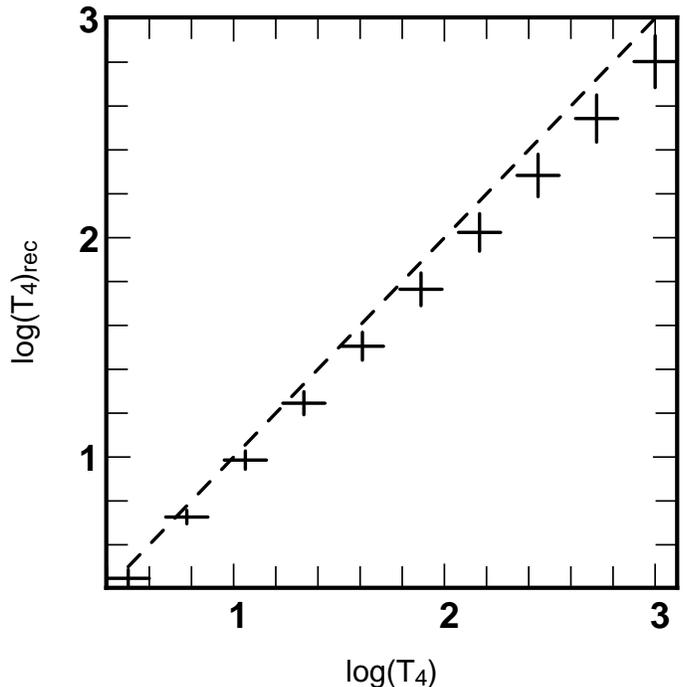}
\caption{recovered $\overline{T}$ versus real $\overline{T}$
from synthetic spectrum described in \Fig{grid_eos_data} 
($T_{4}=\overline{T}/10^4$). The  value of $\beta$ is assumed to be known.
$\overline{T}$ is well recovered in the range  5\,000 K - 30\,000 K,
when the data span a wide range of densities.
The error bars correspond to $\chi^2=1\pm\sqrt{2/n}$, where n is the number of pixels.
For high $\overline{T}$, there is a small bias to lower $\overline{T}$.
}
\label{f:grid_eos_result}
\end{figure}

In  the  presence  of noise,  the  actual  numerical  inversion of  a  given
underlying field, $\rho(x)$, remains a somewhat degenerate problem for small  values of
$\beta  $ since the
${{\chi }^2}  $ {function} {will depend} {weakly}  {on} ${\big(\sigma_{\delta} ,{\overline
T} \big)  }$ near the  minimum (the minimum  is "flat").
For instance, a relative error of $5\%$ on $\sigma_{2}$ and 
$\sigma_{4}$ leads to a relative error of $20\%$ for ${\overline T}$.
We can use \Eq{ss2} to estimate what  range of $\rho$ we need to consider to
better constraint  the temperature.   A remaining issue is 
 how does the inversion deal with lines which have intrinsic kurtosis ? 
It turns out  that statistically (ie averaged over a few such lines) we have
some insight of what the shape of the underlying density profile should be,  
since
the relative distortions induced by $\beta $ are the same for all lines. 
To demonstrate this,
let us investigate how our non-parametric procedure (\Sec{eos})
recovers the temperature, $\overline{T}$,
for a large range  of temperature (5\,000 K to 30\,000 K)
 when dealing with simulated profiles (provided the simulated data span a sufficiently large range of
densities).
\parn
For this purpose, we synthetize a \los\ of  H~{\sc i} clouds 
whose width and height span the range $40<\ b\ ({\rm km\ s}^{-1})\ <200$ 
and $3<\ \rho_{\rm max}\ <10$. This \los\ is plotted in the lower panel of 
\Fig{grid_eos_data}. We generate the corresponding spectrum
 for ten values of $\overline{T}$ between 5\,000 K and  30\,000 K
while  $\beta$ is kept constant and equal to $0.25$. The spectrum corresponding to $\overline{T}=$15\,000 K
is plotted in the upper panel of \Fig{grid_eos_data}. 
The recovered temperature $\overline{T}$ corresponding to a $\chi^2=1$ fit is
plotted versus the real  $\overline{T}$ in  \Fig{grid_eos_result}.
For a non-parametric model, a $\chi^2=1\pm \sqrt{2/N}$ fit
does correspond to the best estimate for the model.  
 The temperature at mean density is better
recovered for low $\overline{T}$. But the bias towards  lower
temperature remains  small up to 30\,000 K.
This  may be due to 
the increasing distance between the true density
and the prior ($D_0$).

    \section{The inversion technique}

We aim to invert \Eq{eqfun}, i.e. reconstruct the density field 
$\rho_{\rm DM}$.  To  that end, we take  a {\em model},
$g$,   which  basically  relates  the  Doppler
parameter $b$ and  the gas density $n_{\rm HI}$ to  the dark matter density,
$\rho_{\rm DM}$ and the parameters of the temperature-density relation.
   The goal  here is  to determine
these unknown fields, $\M{M}=\left(\log(\frac{\rho_{\rm DM}}{\overline{\rho}_{DM}}),\overline{T},\beta\right)$,
by fitting the data,  $\M{D} \equiv \tau(w_{i})$, 
i.e. the optical depth along the \los.
\\
Since the  corresponding inversion problem is under-determined, we use the Bayesian technique described
in Pichon \& al (2001).  In order to achieve regularization,  we require a
prior guess for  the parameters, $\M{M}$, 
or in statistical  terms, their probability
distribution function, $f_{\rm prior}(\M{M})$.
Using Bayes' theorem, the conditional  probability density
$f_{\rm post}(\M{M}|\M{D})$  for the realization $\M{M}$  given the observed
data $\M{D}$ then writes:
\begin{equation}
f_{\rm post}(\M{M}|\M{D})={\cal L}(\M{D}|\M{M})f_{\rm prior}(\M{M}) \,,
\end{equation}
where $\cal L$ is the likelihood function of the data given the model.
If we assume that both functions  $\cal L$ and $f_{\rm prior}$ are Gaussian,
we can write:
\begin{eqnarray}
f_{\rm post}(\M{M}|\M{D})&=&{\cal A}\exp[-\frac{1}{2}(\M{D}-g(\M{M}))^{\perp}
\cdot\M{C}_d^{-1}\cdot(\M{D}-g(\M{M}))\nonumber\\
& &\ -\frac{1}{2}(\M{M}-\M{M}_0)^{\perp}\cdot\M{C}_0^{-1}\cdot(\M{M}-\M{M}_0)] 
\,, 
\EQN{fpost}
\end{eqnarray}
with   $\M{C}_d$   and   $\M{C}_0$   being   respectively   the   covariance
``matrix''
of  the noise  and of  the prior  guess for  the
parameters, $\M{M}_0$.   $\cal A$ is a normalization  constant.  The superscript,
$\perp$, stands for transposition. 
In a nutshell, the minimum of the argument of the
exponential  in  \Eq{fpost}  is found using an iterative procedure:
\begin{eqnarray}
{\M{M}}_{[k+1]}&=&\M{M}_0 + \M{C}_0  \cdot \T{ \M{G}_{[k]} } \cdot
(\M{C}_d + \M{G}_{[k]} \cdot \M{C}_0 \cdot \T{\M{G}_{[k]}})^{-1} \cdot \nonumber\\
&& \ \ \ \ \ \ \ (\M{D}       +     \M{G}_{[k]}          \cdot     ({\M{M}}_{[k]}
-\M{M}_0)-g({\M{M}}_{[k]})) \,,
\EQN{sp2}
\end{eqnarray}
where subscript  $[k]$ refers  to the iteration  order, while
 $\M{G}$ is the matrix (or more rigorously, the functional operator) of
partial  derivatives   of  the  model  $g({\M{M}})$  with   respect  to  the
parameters.  In this  scheme the
minimum corresponds to $\M{M}_{[\infty]}$ and in practice is
found via a convergence criterion  on the relative changes between iteration
$[k]$  and $[k+1]$.

\parn Our model reads, \Eq{sp1},
\begin{eqnarray}
  g_{i}(p)=A(\overline{z})c_1\int_{-\infty}^{+\infty}
(\exp[p(x)])^{\alpha-\beta}\exp\left(-           
c_2            \frac{(w_{i}-
x)^2}{(\exp[p(x)])^{2\beta}} \right) \d x\, ,\nonumber\ 
\end{eqnarray}
where $p=\log(\frac{\rho_{\rm DM}}
{\bar{\rho}_{\rm DM}})$.  The prior is defined as 
$\M{M}_{\rm 0}=(p_{0}=0)$. The treatment of the temperature-density relation
is different and it is described in \Sec{eos}.
Since  the
model,  $\M{M}\equiv  p(x)$  is  a  continuous field,  we  need  to
interpret \Eq{sp2} in terms of convolutions, and functional derivatives.  In
particular  the  matrix   of  partial  functional  (Fr\'echet)  derivatives,
$\M{G}$, has the following kernel:
\begin{eqnarray}
(\M{G})_{i}(x)   \equiv   \left(\frac{\partial   g_{i}}{\partial
p}\right)(x)=    A(\overline{z})    c_1   D_0^{\alpha-\beta}(x)
\exp\left[(\alpha-\beta)p(x)\right]                 B_{i}(x)\nonumber\, ,\nonumber\
\end{eqnarray}
with
\begin{eqnarray}
B_{i}(x)&=&  ( (\alpha-\beta)
+ c_2 2 \beta (w_{i}-x)^2 \exp\left[-2\beta p(x)\right] )\cdot \nonumber\\ 
&&\ \ \ \ \  \exp\left(-c_2 \frac{(w_{i}-x)^2}
{(
\exp\left[p(x)\right])^{2\beta}} \right) \, .\nonumber\
\end{eqnarray}

\end{document}